\documentclass[amsmath,amssymb,amsfonts,aps,pre,preprint,superscriptaddress,bibnotes,showpacs,showkeys,longbibliography]{revtex4-1}
\usepackage{mathtools}
\usepackage{epsfig}
\usepackage[english]{babel}
\usepackage{color}
\usepackage{subcaption}
\usepackage{hyperref}
\usepackage{lipsum}
\usepackage{xcolor}
\usepackage{amsthm}

\newcommand*{\lop}{\mathcal{L}_{t,x}\,}
\newcommand*{\lopt}{\mathcal{L}_{t}\,}
\newcommand*{\nlop}{\mathcal{N}_{t,x}\,}
\newcommand*{\rop}{\mathcal{R}_{x}\,}
\DeclareMathOperator\erf{erf}

\AtBeginDocument{%
    \newwrite\bibnotes
    \def\bibnotesext{Notes.bib}
    \immediate\openout\bibnotes=\jobname\bibnotesext
    \immediate\write\bibnotes{@CONTROL{REVTEX41Control}}
    \immediate\write\bibnotes{@CONTROL{%
    apsrev41Control,author="08",editor="1",pages="1",title="0",year="1"}}
     \if@filesw
     \immediate\write\@auxout{\string\citation{apsrev41Control}}%
    \fi
}%

\newcommand{\me}{\mathrm{e}}

\begin{document}
\title{The BLUES function method applied to partial differential equations and analytic approximants for interface growth under shear}

\author{Jonas Berx}
\affiliation{Institute for  Theoretical Physics, KU Leuven, B-3001 Leuven, Belgium}

\author{Joseph O. Indekeu}
\affiliation{Institute for  Theoretical Physics, KU Leuven, B-3001 Leuven, Belgium}

\date{\today}

\begin{abstract}
An iteration sequence based on the BLUES (Beyond Linear Use of Equation Superposition) function method is presented for calculating analytic approximants to solutions of nonlinear partial differential equations. This extends previous work using this method for nonlinear ordinary differential equations with an external source term. Now, the initial condition plays the role of the source. The method is tested on three examples: a reaction-diffusion-convection equation, the porous medium equation with growth or decay and the nonlinear Black-Scholes equation. A comparison is made with three other methods: the Adomian decomposition method (ADM), the variational iteration method (VIM) and the variational iteration method with Green function (GVIM). As a physical application, a deterministic differential equation is proposed for interface growth under shear, combining Burgers and Kardar-Parisi-Zhang nonlinearities. Thermal noise is neglected. This model is studied with Gaussian and space-periodic initial conditions. A detailed Fourier analysis is performed and the analytic coefficients are compared with those of ADM, VIM, GVIM and standard perturbation theory. The BLUES method turns out to be a worthwhile alternative to the other methods. The advantages that it offers ensue from the freedom of choosing judiciously the  linear part, with associated Green function, and the residual containing the nonlinear part of the differential operator at hand. \\
\end{abstract}

\maketitle

\section{Introduction}\label{sec:intro}
It is a challenge, in exact sciences and theoretical physics in particular, to obtain useful analytical approximations to solutions of nonlinear differential equations (DEs). In this context the Adomian decomposition method (ADM), the homotopy analysis method (HAM) or perturbative techniques such as the soliton perturbation theory have proven useful \cite{adomian,ham,karpman1977,Keener}. In two recent papers \cite{Berx,Berx_2020}, we demonstrated how the practice of Green functions can be usefully extended to nonlinear {\em ordinary} differential equations (ODEs) that are inhomogeneous, featuring a source or sink, by effectively using the superposition principle beyond the linear domain. In the present paper, we extend the approach to nonlinear {\em partial} differential equations (PDEs) and present an application to the physics of interface growth in a soft condensed matter system under shear flow.

To situate this development, we briefly recall the history of the BLUES function method.  In \cite{smets} exponential tail solutions of nonlinear reaction-diffusion-convection ODEs describing traveling wave fronts with co-moving sources were studied.
In \cite{BLUES} it was noted that an exponential tail solution may simultaneously solve the nonlinear ODE  and a related linear ODE, both with a co-moving Dirac delta source. This led to an analytic method that uses the Green function beyond the linear domain, named BLUES (``Beyond Linear Use of Equation Superposition"). Next, in \cite{Berx,Berx_2020} it was shown how to develop the method into a non-perturbative and rapidly converging analytic iteration procedure. One may start from a linear DE and freely add  a nonlinearity.   Applications were given to solitary waves, oscillatory waves, nonlinear growth and transport of heat, and the method was extended to fractional ODEs and to sources that need not be co-moving. 

Now, we extend the approach to nonlinear PDEs, e.g., in time $t$ and one space coordinate $x$, which cannot be reduced to ODEs. For PDEs the initial condition serves as the source and no external source must be added. We will compare the BLUES iteration with four other methods: the Adomian decomposition method (ADM) \cite{adomian,ADOMIAN199017}, the variational iteration method (VIM) \cite{HE20073}, the VIM with Green function (GVIM) \cite{KHURI201428},  and straightforward perturbation theory (PT). 

The setup of this work is as follows. In Section \ref{sec:BLUES_method_pde} we extend the BLUES function method to the arena of PDEs in two variables, one of which is time. We restrict our attention in this paper to operators with a first derivative in time. In Section \ref{sec:testcases} we illustrate the method for three simple exactly solvable PDEs and compare the different methods. In Section \ref{sec:general} we set the stage for a physical problem by applying the method to a general power-law convective nonlinearity. Next, in Section \ref{sec:interface} we introduce and study a simple model for the time evolution of a growing fluid interface under shear. In Section \ref{sec:conclusions} we conclude and present an outlook.

\section{The BLUES function method for a nonlinear PDE}\label{sec:BLUES_method_pde}
Here we extend the BLUES iteration method originally developed for ODEs \cite{BLUES,Berx,Berx_2020} to PDEs in time and one space variable. The crucial role of the  extrinsic {\em source (or sink) term} in the context of the ODE will now be taken over, simply, by the intrinsic {\em initial condition} of the solution of the PDE. Consequently, the extension of the method to PDEs entails a conceptual simplification rather than complication, and allows one to increase substantially the range of physics problems that can be tackled.

Let us start from a linear PDE which can be written as an operator $\lop$ acting on a function $u(x,t)$, say a density subject to diffusion, and let us attempt to solve 
\begin{equation}
    \label{eq:linear_operator_delta}
    \lop u(x,t) = 0, \; \mbox{for} \; t >0,
\end{equation}
with initial condition
\begin{equation}
    \label{eq:linear_operator_delta_initial}
     u(x,0) = f(x).
\end{equation}
Since the problem is linear the solution $u(x,t)$ can be written as the convolution $G \ast f$ of the initial condition $f(x)$ with the Green function $G(x,t)$, which satisfies
\begin{equation}
    \label{eq:linear_operator_delta_G}
    \lop G(x,t) = 0, \; \mbox{for} \; t >0,
\end{equation}
with Dirac-delta initial condition
\begin{equation}
    \label{eq:linear_operator_delta_initial_G}
     {\rm lim}_{t\rightarrow 0} \, G(x,t) = \delta (x),
\end{equation}
The solution to the linear problem is the (single-variable) convolution, which reads
\begin{equation}
    \label{eq:linear_operator_delta_u}
    u(x,t) = \int _{\mathbb{R}} dx' \,G(x-x',t) f(x').
\end{equation}

For simplicity we restrict our attention to PDEs that involve only the first derivative w.r.t. to time, specifically $\lop u = u_t + \tilde{\mathcal{L}_{x}}u$, with $u_t \equiv \partial u/\partial t$ and $\tilde{\mathcal{L}_{x}}$ a time-independent linear operator. 
For our purposes, it is convenient to rewrite the PDE by invoking the initial condition
$f (x)$ through the action of a Dirac-delta source in time. The following time and space integral, which is a two-variable convolution $u(x,t) = G \ast f\,\delta$, solves the  rearranged inhomogeneous linear PDE, which is equivalent to the original linear PDE,
\begin{equation}
    \label{eq:rewrite_linear_operator_delta_u}
   \lop u(x,t) =   \lop \int _ {0^-} ^t \, dt' \int _{\mathbb{R}} dx' \,G(x-x',t-t') f(x') \delta (t') = f(x) \delta (t)
\end{equation}

This identity holds by virtue of the fact that $\lop$ contains only a first
derivative w.r.t. time $t$. This derivative generates two terms. The boundary term (the value of the integrand at $t'=t$) exactly produces the right-hand-side of \eqref{eq:rewrite_linear_operator_delta_u}, in view of \eqref{eq:linear_operator_delta_initial_G}. The second term is contained in the action of $\lop$, when it is moved inside the integral over $t'$. That contribution, however, vanishes as one can verify by careful inspection. We conclude that $G \ast f\,\delta$ solves the PDE for all $t > 0$.

The initial condition is retrieved by examining  the limit $t \rightarrow 0$. Firstly, the solution $u(x,t)$ as given by the time and space integral $G \ast f\,\delta$ obviously vanishes for $t<0^-$ by definition, so $u(x, t < 0) = 0$. However, this solution ``jumps" to the initial condition function $f(x)$ at $t = 0^+$ through the action of  $\delta(t')$ and by the fact that the Green function becomes a spatial Dirac-delta in view of \eqref{eq:linear_operator_delta_initial_G}. The space integral then produces $f(x)$. For $t>0$ the solution evolves, in a continuous manner, from this initial condition. 

Using this representation of the PDE, which naturally features an intrinsic source term expressing the initial condition, we can now generalize the BLUES iteration procedure from nonlinear ODEs to nonlinear PDEs. One may add a nonlinearity rather freely to the PDE, while preserving the simple form of the time-dependent part,
\begin{equation}
\nlop u = u_t + \tilde{\mathcal{N}_{x}}u, 
\end{equation}
with $\tilde{\mathcal{N}_{x}}$ a time-independent nonlinear operator,
 and arrive at the nonlinear PDE
\begin{equation}
    \label{eq:nonlinear_PDE_psi}
    \nlop u(x,t) = 0,
\end{equation}
with intitial condition, as before,
\begin{equation}
    \label{eq:linear_operator_delta_initial_2}
     u(x,0) = f(x).
\end{equation}

The BLUES function method now proposes to construct a solution $u(x,t)$ to the equivalent inhomogeneous PDE in the form of a two-variable convolution $u(x,t) = B \ast \phi$, so that
 \begin{equation}
    \label{eq:rewrite_nonlinear_operator_delta_u}
   \nlop u(x,t) =   \nlop \int _ {0^-} ^t \, dt' \int _{\mathbb{R}} dx' \,B(x-x',t-t') \phi(x',t') = f(x) \delta (t).
\end{equation}
Clearly, this PDE coincides with the original nonlinear PDE \eqref{eq:nonlinear_PDE_psi} for $t>0$ and we will shortly examine its behavior at $t=0$. The function $B(x,t)$ is called BLUES function and it is taken to be the Green function of an arbitrary but conveniently chosen linear operator $\lop$ related to $\nlop$. The challenge is to calculate the new associated source $\phi(x,t)$ knowing that $B \ast f \delta$ solves the linear PDE \eqref{eq:rewrite_linear_operator_delta_u} with initial condition $f(x)$ and source term $f \delta$. Note that $\phi(x,t)$ need not be separable and in general it is not.

The initial condition is generated correctly, since, by definition, $ u(x, t < 0) = 0$  and subsequently $u(x, t = 0^+) = f (x)$, provided three conditions are fulfilled. The first is that $\nlop u = 0$, for $u=0$. The second condition is that the associated source $\phi$ decomposes as follows into a separable singular term and a (non-separable) smooth term $\zeta$, which is to be calculated analytically$:$ 
$\phi(x,t) = f (x) \delta(t) + \zeta(x,t)$, with $\int _{0^-} ^{0^+} dt \, \zeta (x,t) =0$. The third condition is that for all finite $x$ the function  $\tilde{\mathcal{N}_{x}}f(x)$ be finite. For nonlinear operators these are not obvious and must be checked.

For this calculation one defines a (time-independent) residual operator $\rop \equiv \lop - \nlop$ and makes use of the implicit identity
\begin{equation}
    \label{eq:phi_identity}
    \nlop (B\ast\phi) = \phi(x,t) - \rop (B\ast\phi) = f(x)\delta(t), 
\end{equation}
which follows directly from the Green function property of $B$ w.r.t. the chosen  linear PDE.

To obtain the solution to the nonlinear PDE \eqref{eq:nonlinear_PDE_psi} with initial condition \eqref{eq:linear_operator_delta_initial_2}, equation \eqref{eq:phi_identity} can be rewritten and iterated,
\begin{equation}
    \label{eq:phi_identity_re}
  \phi(x,t)  = f(x)\delta(t) +  \rop (B\ast\phi), 
\end{equation}
 in order to calculate an approximation in the form of a sequence in powers of the residual $\rop$. In zeroth iteration, 
\begin{equation}
    \label{eq:zeroth_order}
    \phi^{(0)}(x,t) = f(x) \delta (t),
\end{equation}
and in $n$th iteration ($n\geq1$), 
\begin{equation}
    \label{eq:nth_order}
    \phi^{(n)}(x,t) = f(x) \delta (t) + \rop (B \ast \phi^{(n-1)}).
\end{equation}
Consequently, the $n$th analytical approximant to the solution of the nonlinear PDE is found through the two-variable convolution
\begin{equation}
    \label{eq:nth_order_u}
    u^{(n)}(x,t) = B \ast \phi^{(n)} = u^{(0)}(x,t) + (B\ast \rop u^{(n-1)})(x,t).
\end{equation}

\section{Test cases for the method}
\label{sec:testcases}
\subsection{Reaction-diffusion-convection equation}
\label{sec:ramos}
Let us start with a simple example, in which the convolutions are all of single variable type. Unless otherwise stated the functions, variables and parameters are reduced (dimensionless).
Consider the nonlinear reaction-diffusion-convection PDE \cite{ramos} which can be used to describe, e.g., the propagation of a chemical of density $u$ through the combined mechanisms of diffusion, nonlinear convection and reaction, 
\begin{equation}
    \label{eq:ramos_pde}
    \begin{split}
    \nlop u &= u_t - u_{xx} + u u_x +u (u+2) = 0
    \end{split}
\end{equation}
defined on $(x,t) \in \mathbb{R} \times [0,\infty)$ with an exponential initial condition, i.e.,
\begin{equation}
    \label{eq:ramos_ic}
    u(x,0) = f(x) = \me^{-x}\, .
\end{equation}
This unbounded initial condition is rather unphysical but will serve as an ideal testbed for the comparison of the different approximation methods, as in this case a simple exact solution of \eqref{eq:ramos_pde} can be found. We will now consider the methods mentioned in Section \ref{sec:intro} and compare their results. 
The ADM and VIM both produce the following sequence of approximants,
\begin{equation}
    \label{eq:ramos_approximants}
    \begin{split}
        u^{(0)}(x,t) &= \me^{-x} \\
        u^{(1)}(x,t) &= \me^{-x}(1-t)\\
        u^{(2)}(x,t) &= \me^{-x}(1-t+\frac{t^2}{2!})\\
                &\vdots\\
        u^{(n)}(x,t) &= \me^{-x}\sum_{i=0}^n \frac{(-t)^i}{i!}\\
    \end{split}
\end{equation}
which converges slowly to the exact solution
\begin{equation}
    \label{eq:ramos_exact}
    u(x,t) = \lim_{n\rightarrow\infty}u^{(n)}(x,t) = \me^{-(x+t)}\, .
\end{equation}
Note that the sequence \eqref{eq:ramos_approximants} is the Taylor series of the temporal part of the exact solution expanded about $t=0$ and hence only useful for $t < {\cal O}(1)$. The GVIM calculations result in a different sequence of approximants,
\begin{equation}
    \label{eq:ramos_VIM_green_approximants}
    \begin{split}
        u^{(0)}(x,t) &= \me^{-x} \\
        u^{(1)}(x,t) &= \frac{\me^{-x}}{2}+\frac{\me^{-2t-x}}{2}\\
        u^{(2)}(x,t) &= \frac{\me^{-x}}{4}+\frac{3\me^{-2t-x}}{4}+\frac{\me^{-2t-x}}{2}t\\
        u^{(3)}(x,t) &= \frac{\me^{-x}}{8}+\frac{7\me^{-2t-x}}{8}+\frac{3\me^{-2t-x}}{4}t +\frac{\me^{-2t-x}}{4}t^2\\
        &\vdots\\
        u^{(n)}(x,t) &= \frac{\me^{-x}}{2^n} + \me^{-2t-x}\sum_{i=0}^{n-1} \frac{2^{n-i}-1}{2^{n-i}i!}t^i\\
    \end{split}
\end{equation}
which converges to the exact solution \eqref{eq:ramos_exact} for $n\rightarrow\infty$ as well.

We now turn to the BLUES function method, and follow the scheme outlined in Section \ref{sec:BLUES_method_pde}. 
First, the PDE \eqref{eq:ramos_pde} with initial condition $f(x)$ is rewritten as follows
\begin{equation}
    \label{eq:ramos_blues_pde}
    \begin{split}
    \nlop u &= u_t - u_{xx} + u u_x +u (u+2) = f(x)\delta(t)
    \end{split}
\end{equation}
defined on $(x,t) \in \mathbb{R} \times [0,\infty)$ and the initial condition $u(x,0) = f(x) = \me^{-x}$ has been converted to a source term by multiplication with a Dirac-delta function in the temporal coordinate. Choosing the linear operator simple and without spatial derivatives, one can define the associated linear PDE with source $\psi(x,t) \equiv f(x)\delta(t)$ as follows,
\begin{equation}
    \label{eq:ramos_blues_linear_operator}
    \lopt u = u_t + 2u = \psi(x,t),
\end{equation}
which is solved by $u(x,t) = f(x) G(t)$, with $G(t)$ the Green function for $ \lopt$.  Note that we  omitted the linear term $u_{xx}$ from the linear part $\lop$ of the operator $\nlop$. This judicious choice, which is a distinct feature of the BLUES strategy, not only simplifies the calculations but also considerably improves the convergence.

We obtain a step function with exponential tail,
\begin{equation}
    \label{eq:ramos_blues_greenfunction}
    G(t) = \Theta(t) \me^{-2t},
\end{equation}
and the solution $U(t)$ for the linear problem with arbitrary source $\psi (t)$, for $t>0$, is
\begin{equation}
    \label{eq:ramos_blues_greenfunction_arb}
    U(t) = G \ast \psi = \int_\mathbb{R} ds \,G(t-s) \psi (s) = \int_{0^-}^t ds \, G(t-s) \psi(s),
\end{equation}
since $G(\tau <0) = 0$ and $s >0$.

We next define the residual operator $\rop$ as the difference between the linear and the nonlinear operator, i.e., $\rop  = \lopt -\nlop $, so
\begin{equation}
    \label{eq:ramos_blues_residual}
    \rop u = u_{xx} - uu_x -u^2
\end{equation}
and set up the iteration sequence based on \eqref{eq:nth_order} and \eqref{eq:nth_order_u} for the solution to \eqref{eq:ramos_blues_pde},
\begin{equation}
    \label{eq:ramos_blues_procedure}
    \begin{split}
    u^{(n+1)}(x,t) &= u^{(0)}(x,t) + (B\ast \rop u^{(n)})(x,t)\\
    &= u^{(0)}(x,t) + \int\limits_{0^-}^t \mathrm{d}s\, G(t-s)\rop u^{(n)}(x,s)\\
    &= u^{(0)}(x,t) + \int\limits_{0^-}^t \mathrm{d}s\, G(t-s)\left[u^{(n)}_{xx}(x,s) - u^{(n)}(x,s)u^{(n)}_x(x,s) -(u^{(n)}(x,s))^2\right],
    \end{split}
\end{equation}
where the BLUES function $B(\tau)$ is  the Green function $G(\tau)$ of \eqref{eq:ramos_blues_greenfunction} for the chosen linear operator $\lopt$, whose action is given in \eqref{eq:ramos_blues_linear_operator}. 
The zeroth  approximant is the convolution of the BLUES function with the source $\psi(x,t)$, 
\begin{equation}
    \label{eq:ramos_blues_zerothorder}
    \begin{split}
    u^{(0)}(x,t) &= \int\limits_{0^-}^t G(t-s)\psi(x,s)\mathrm{d}s = \me^{-2t -x}\, .
    \end{split}
\end{equation}
Iterating through the procedure \eqref{eq:ramos_blues_procedure}, one finds the following sequence of approximants
\begin{equation}
    \label{eq:ramos_blues_approximants}
    \begin{split}
    u^{(0)}(x,t) &= \me^{-2t-x}\\
    u^{(1)}(x,t) &= \me^{-2t-x} (1 + t)\\
    u^{(2)}(x,t) &= \me^{-2t-x}(1 + t + \frac{t^2}{2!})\\
        &\vdots\\
    u^{(n)}(x,t) &= \me^{-2t-x} \sum_{i=0}^n\frac{t^i}{i!}\, ,
    \end{split}
\end{equation}
which converges to the exact solution \eqref{eq:ramos_exact} for $n\rightarrow\infty$. Note that each approximant is bounded and useful for all $t$ by virtue of the overall factor $\me^{-2t}$.

We can now compare the results of the three different methods. Since all three methods converge to the known exact solution \eqref{eq:ramos_exact}, one can define an error function $E^{(n)}(x,t)$  as the absolute value of the difference between the $n$th approximant and the exact solution $u_{ex}$, 
\begin{equation}
    \label{eq:ramos_blues_error}
    E^{(n)}(x,t) = |u_{ex}(x,t) - u^{(n)}(x,t)|
\end{equation}

\begin{figure}[!ht]
    \centering
    \begin{subfigure}{0.48\textwidth}
        \includegraphics[width=\linewidth]{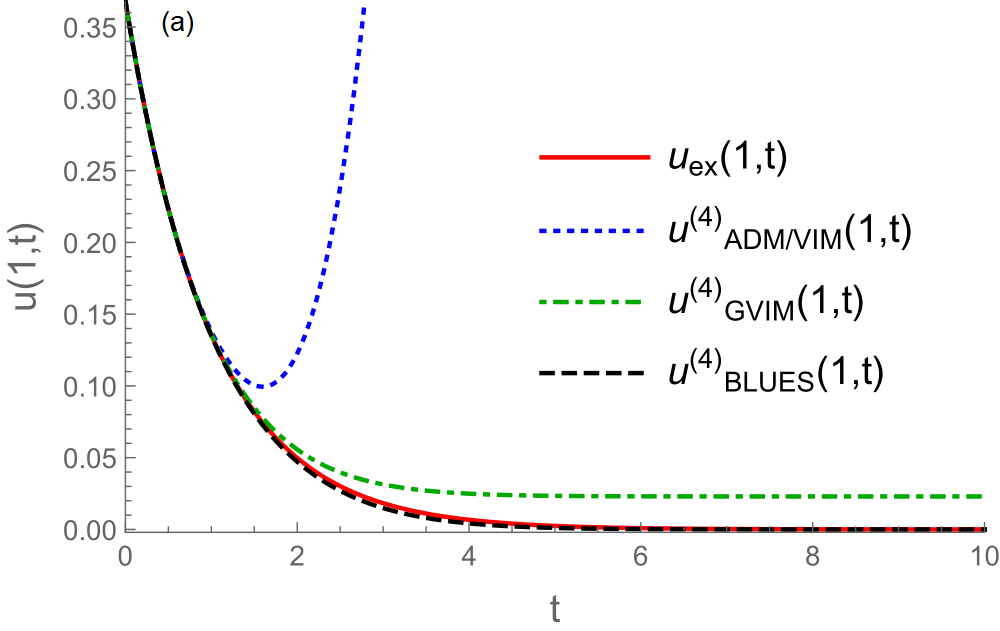}
    \end{subfigure}
    \begin{subfigure}{0.48\textwidth}
        \includegraphics[width=\linewidth]{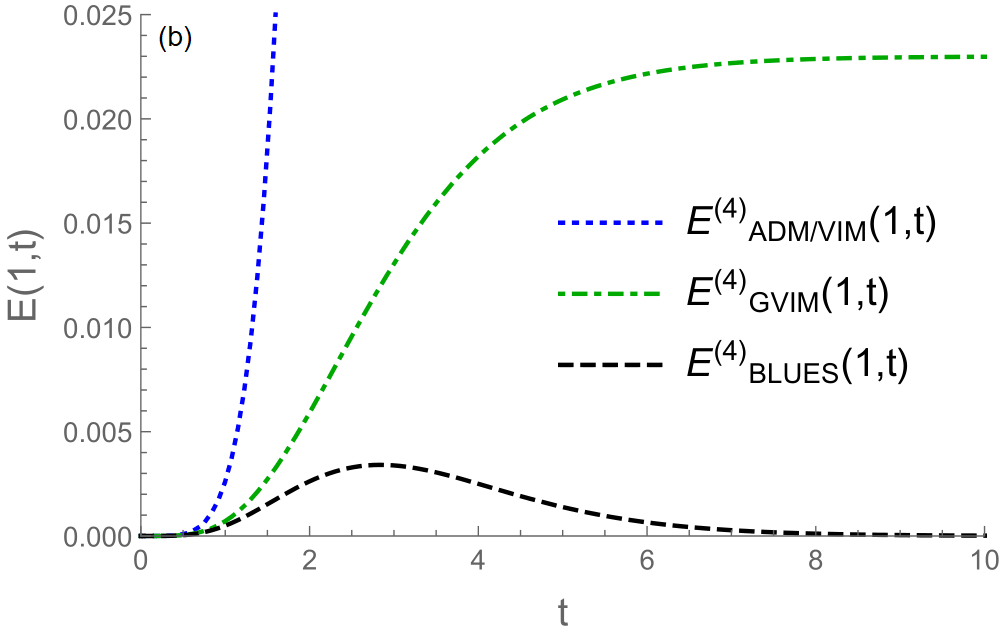}
    \end{subfigure}
    \caption{Reaction-diffusion-convection equation. \textbf{(a)} The approximants $u^{(4)}(x=1,t)$ and the exact solution (red, full line) \eqref{eq:ramos_exact}. \textbf{(b)} The errors $E^{(4)}(x=1,t)$ for the different methods: ADM and VIM \eqref{eq:ramos_approximants} (blue, dotted line), GVIM \eqref{eq:ramos_VIM_green_approximants} (green, dot-dashed line) and BLUES \eqref{eq:ramos_blues_approximants} (black, dashed line).}
    \label{fig:ramos_comparison_error}
\end{figure}

In Fig. \ref{fig:ramos_comparison_error}, the approximants $u^{(n)}(x,t)$ and the errors $E^{(n)}(x,t)$ for the different methods are shown for $n = 4$ and fixed position $x=1$. One can observe that the error in ADM and VIM   becomes very large for values of $t\gg1$, indicating that the approximants diverge for large $t$, as expected. The error in the GVIM, however, saturates at a finite value which can be calculated for all values of $x$ as
\begin{equation}
    \lim_{t\rightarrow\infty}E^{(n)}_{\rm GVIM}(x,t) = \frac{\me^{-x}}{2^n}\, ,
\end{equation}
which for $n = 4$ and $x = 1$ results in $(16e)^{-1}$. Note that the errors for both the ADM and VIM and for the GVIM are monotonically increasing in time and hence the approximations decrease in accuracy for large values of $t$. In contrast, for the BLUES function method the error vanishes in the limit $t\rightarrow\infty$ and this method provides the fastest convergence for all  $t>0$. The reason for this improved performance is that the {\em choice of the linear operator part} in the BLUES function method is free and can be tailored so as to render all the approximants well bounded for all times.

\subsection{Porous medium equation with growth or decay}
\label{sec:porous_medium}

The second example is in the realm of fluid mechanics: the nonlinear porous medium equation \cite{vazquez} with linear growth or decay, 
\begin{equation}
    \label{eq:porous_PDE}
    w_t - \Delta(w^m) - \beta w = 0,
\end{equation}
with $m>1$ and $\beta\in\mathbb{R}$. We consider a density $w(x,t)$ in one space dimension with initial condition $w(x,0) = f(x)= x$. Unless otherwise stated the functions, variables and parameters are reduced (dimensionless). We will only consider a quadratic nonlinearity, $m=2$, which allows us to write \eqref{eq:porous_PDE} as follows
\begin{equation}
    \label{eq:porous_PDE_2}
    w_t - 2 w w_{xx} -2 w_x^2 - \beta w = 0\, .
\end{equation}
The components of the solution generated by the ADM are
\begin{equation}
    \label{eq:porous_approximants_ADM}
    \begin{split}
        w_0(x,t) &= x\\
        w_1(x,t) &= 2 t + \beta x t \\
        w_2(x,t) &= 3 \beta t^2 + \beta^2 x \frac{t^2}{2} \\
w_3(x,t) &= \frac{7\beta^2t^3}{3}+ \beta^3x \frac{t^3}{6}\\
                &\vdots\\
        w_i(x,t) &= \frac{2 (2^i -1) \beta^{i-1} t^i }{i!} + x \frac{\beta^i t^i}{i!}
    \end{split}
\end{equation}
for $i\geq1$. The $n$th-order approximant is the partial sum of the component functions $w_i$, 
\begin{equation}
    w_{\rm ADM}^{(n)}(x,t) =  \sum\limits_{i=0}^n w_i(x,t)\, ,
\end{equation}
and in the limit $n \rightarrow \infty$ this converges to the exact solution 
\begin{equation}
    \label{eq:porous_exact}
    w(x,t) = \lim_{n\rightarrow\infty}w^{(n)}(x,t) = (x-\frac{2}{\beta})\me^{\beta t}+\frac{2}{\beta}\me^{2\beta t}\, ,
\end{equation}
where the sign of $\beta$ indicates whether there is growth or decay. 
Note that  the ADM generates term by term the exact coefficients of the powers of $t$ in the Taylor expansion in time of the solution. 

The VIM produces the following sequence of approximants to the solution of \eqref{eq:porous_PDE_2},
\begin{equation}
    \label{eq:porous_approximants}
    \begin{split}
        w^{(0)}(x,t) &= x\\
        w^{(1)}(x,t) &= 2 t +x + \beta xt \\
        w^{(2)}(x,t) &= 2 t + 3 \beta t^2 + \frac{2\beta^2 t^3}{3}+ x\left(1 + \beta t + \frac{\beta^2 t^2}{2}\right)\\
        w^{(3)}(x,t) &= 2 t + 3 \beta t^2 + \frac{7\beta^2 t^3}{3} + \frac{2\beta^3 t^4}{3} +\frac{\beta^4 t^5}{10} + x\left(1 +\beta  t + \frac{\beta^2 t^2}{2} +\frac{\beta^3 t^3}{6}\right)\\
        &\vdots,\\
    \end{split}
\end{equation}
which also converges to the exact solution \eqref{eq:porous_exact}. 
Note that VIM and ADM produce different results. The VIM does not immediately give the exact coefficients but recursively adjusts them until they saturate at the exact value. 

Next, the GVIM produces the sequence
\begin{equation}
    \label{eq:porous_VIM_green_approximants}
    \begin{split}
        w^{(0)}(x,t) &= x \\
        w^{(1)}(x,t) &= -\frac{2}{\beta }+\frac{2 e^{\beta  t}}{\beta }+x e^{\beta  t}\\
        w^{(2)}(x,t) &= x\me^{\beta t}-\frac{2}{\beta}\me^{\beta t}+\frac{2}{\beta}\me^{2\beta t}\\
                &\vdots\\
        w^{(n)}(x,t) &=x\me^{\beta t}-\frac{2}{\beta}\me^{\beta t}+\frac{2}{\beta}\me^{2\beta t}
    \end{split}
\end{equation}
For $n \geq 2$, the approximants \eqref{eq:porous_VIM_green_approximants} are invariable.  The GVIM  in this case produces the exact solution \eqref{eq:porous_exact} already in the second iteration and contributions from higher iterations are zero. 

\begin{figure}[!ht]
    \centering
    \begin{subfigure}{0.48\textwidth}
        \includegraphics[width=\linewidth]{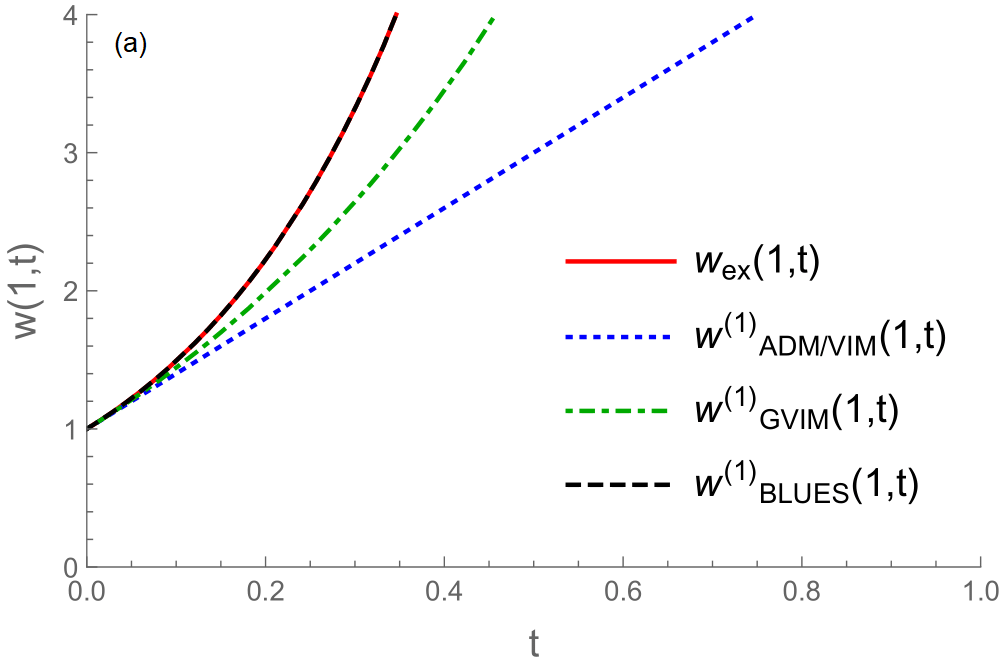}
    \end{subfigure}
    \begin{subfigure}{0.48\textwidth}
        \includegraphics[width=\linewidth]{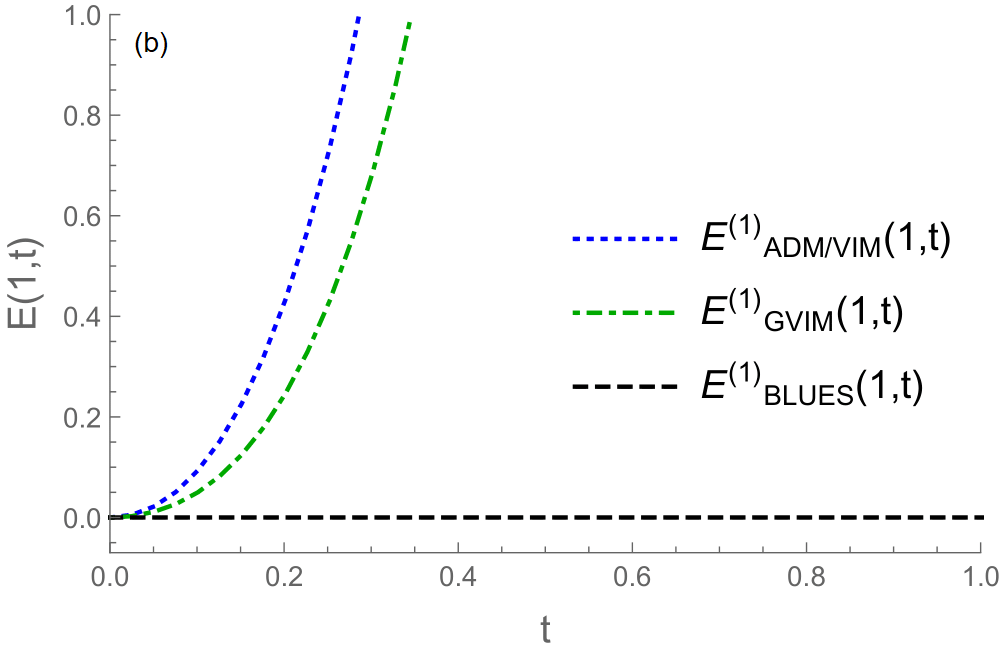}
    \end{subfigure}
    \caption{Porous medium equation. \textbf{(a)} Exact solution $w_{\rm ex}$ (red, full line) and approximants $w^{(1)}$ in first iteration or first order. Note that the ADM and VIM give identical results at this order $n=1$. The BLUES approximant is exact. \textbf{(b)} Difference $E^{(1)}$ between the exact solution \eqref{eq:porous_exact} and the approximant of order $n =1$ for the different methods: ADM and VIM  \eqref{eq:porous_approximants} (blue, dotted line), GVIM \eqref{eq:porous_VIM_green_approximants} (green, dot-dashed line), BLUES \eqref{eq:porous_blues_approximants} (black, dashed line). The parameter $\beta$ takes the value 2. The position in space is fixed at $x = 1$.}
    \label{fig:porous_comparison_error}
\end{figure}

We now turn to the BLUES function method. The PDE \eqref{eq:porous_PDE} with initial condition $w(x,0) = f(x)$ can  be rewritten as a nonlinear PDE with a source $\psi(x,t) = f(x)\delta(t)$,
\begin{equation}
    \label{eq:porous_blues_pde}
    \begin{split}
    \nlop w &= w_t - \left(w^m\right)_{xx} - \beta w = \psi(x,t)
    \end{split}
\end{equation}
defined on $(x,t)\in \mathbb{R} \times [0,\infty)$. Choosing the linear operator to be of the same form as the successful one used in the previous section, one can define the associated linear PDE with the same source term,
\begin{equation}
    \label{eq:porous_blues_linear_operator}
    \lopt w = w_t - \beta w = \psi(x,t)
\end{equation}
and we recall the Green function for this linear operator,
\begin{equation}
    \label{eq:porous_blues_greenfunction}
    G(t) = \Theta (t) \me^{\beta t}
\end{equation}
Note that in this case the linear operator is chosen by simply dropping (only) the nonlinear term in $\nlop$.
We now obtain the residual operator $\rop $, which acts as follows on the function $w$, 
\begin{equation}
    \label{eq:porous_blues_residual}
    \rop w = (w^m)_{xx}
\end{equation}
and set up the iteration sequence for the solution to \eqref{eq:porous_blues_pde}
\begin{equation}
    \label{eq:porous_blues_procedure}
    \begin{split}
    w^{(n+1)}(x,t) &= w^{(0)}(x,t) + (B\ast \rop w^{(n)})(x,t)\\
    &= w^{(0)}(x,t) + \int\limits_{0^-}^t \mathrm{d}s\, G(t-s)\rop w^{(n)}(x,s)\\
    &= w^{(0)}(x,t) + \int\limits_{0^-}^t \mathrm{d}s \, G(t-s)  (w^{(n)})^m_{xx}(x,s),
    \end{split}
\end{equation}
where the BLUES function $B(\tau)$ is the Green function $G(\tau)$ of \eqref{eq:porous_blues_greenfunction} for the chosen linear operator $\lopt$, whose action is given in \eqref{eq:porous_blues_linear_operator}. 
The zeroth approximant is the convolution of the BLUES function  and the source $\psi(x,t)$, i.e.,
\begin{equation}
    \label{eq:porous_blues_zerothorder}
    \begin{split}
    w^{(0)}(x,t) &= \int\limits_{0^-}^t \mathrm{d}s\, G(t- s)\psi(x,s) = x\,\me^{\beta t}\, .
    \end{split}
\end{equation}
Iterating further according to the procedure \eqref{eq:porous_blues_procedure}, one finds the following sequence of approximants for $m=2$
\begin{equation}
    \label{eq:porous_blues_approximants}
    \begin{split}
    w^{(0)}(x,t) &= x\me^{\beta t}\\
    w^{(1)}(x,t) &=x\me^{\beta t}-\frac{2}{\beta}\me^{\beta t}+\frac{2}{\beta}\me^{2\beta t}\\
        &\vdots\\
    w^{(n)}(x,t) &=x\me^{\beta t}-\frac{2}{\beta}\me^{\beta t}+\frac{2}{\beta}\me^{2\beta t}\, ,
    \end{split}
\end{equation}
which, remarkably, produces the exact solution \eqref{eq:porous_exact} to \eqref{eq:porous_PDE} already in the first iteration. Higher iterations remain at this ``fixed point". In Fig.\ref{fig:porous_comparison_error} we compare the results from each of the above methods and also compare their errors, at the level of this first iteration.

\subsection{Nonlinear Black-Scholes equation}\label{sec:black_scholes}

For the following example, let us look at the field of economics. Unless otherwise stated the functions, variables and parameters are reduced (dimensionless). The Black-Scholes equation describes the value $V(S,\tau)$ of an option for some underlying asset price $S\in[0,\infty)$ over a period $\tau\in [0,T]$, with $T$ the time of maturity, that is, the last moment on which an option can be exercised. After expiration or maturity, the option contract will cease to exist and the buyer cannot exercise their right to buy or sell. The underlying asset price $S$ is a stochastic variable and follows a geometric Brownian motion. In \cite{Esekon_2013}, the authors consider a nonlinear Black-Scholes PDE for $V(S,\tau)$, which assumes that the market is incomplete through the combined feedback effects of illiquid markets and large trader effects. In this PDE $S$ is treated as a continuous variable, which we name $s$, and $s$ and $\tau$ are treated as independent variables. This PDE is the following,
\begin{equation}
    \label{eq:pde_black_scholes}
    u_t + \frac{\sigma^2 s^2}{2} u_{ss} \left(1+2\rho \,s \,u_{ss}\right) + r s \, u_s - r u = 0\, ,
\end{equation}
with $t$ the time until expiry, $t=T-\tau$, $u$ the value function, $u(s,t) \equiv V(S,\tau)$, $\sigma$ the volatility, $r$ the risk-free interest rate. The constant $\rho$ is a measure of the liquidity of the market. In order to ensure that feedback effects from hedging generate so-called \emph{volatility smiles}, one has to choose this liquidity parameter to be negative \cite{Platen1998,Frey2002}. We consider the initial condition $u(s,0) = f(s) = s - \sqrt{s S_0}/\rho - S_0/(4\rho)$, where $S_0 \equiv S(\tau=0)$ is the starting price of the asset. 

In \cite{Gonzalez-Gaxiola}, the authors study the solution of \eqref{eq:pde_black_scholes} by means of the ADM. This gives the following sequence of component functions of the solution,
\begin{equation}
    \label{eq:black_scholes_ADM_approximants}
    \begin{split}
        u_0(s,t) &= s - \frac{\sqrt{s S_0}}{\rho} - \frac{S_0}{4\rho}\\
        u_1(s,t) &= -\frac{(4r+\sigma^2)}{8\rho}\left(\frac{S_0}{2} + \sqrt{sS_0}\right)t\\
        u_2(s,t) &= -\frac{(4r+\sigma^2)^2}{128\rho}\left(S_0 + \sqrt{sS_0}\right)t^2\\
        &\vdots\\
    \end{split}
\end{equation}
The solution is the sum of all the component functions $u_i(s,t)$, 
\begin{equation}
    u_{\rm ADM}(s,t) = \sum\limits_{i=0}^\infty u_i(s,t)\, .
\end{equation}
This claim can easily be verified by noticing that the component functions $u_i(s,t)$ are the coefficients of the Taylor series of the exact solution \cite{Esekon_2013},
\begin{equation}
       \label{eq:black-scholes_exact}
       u(s,t) = s - \frac{\sqrt{S_0}}{\rho}\left(\sqrt{s}\,\me^{(r+\frac{\sigma^2}{4})t/2} + \frac{\sqrt{S_0}}{4}\,\me^{(r+\frac{\sigma^2}{4})t}\right)\, .
    \end{equation}

The VIM produces the following sequence of approximants to the solution of \eqref{eq:pde_black_scholes},
   \begin{equation}
        \label{eq:black_scholes_vim_approximants}
        \begin{split}
            u^{(0)}(s,t) &= s - \frac{\sqrt{s S_0}}{\rho} - \frac{S_0}{4\rho}\\
            u^{(1)}(s,t) &= s - \frac{\sqrt{s S_0}}{\rho} - \frac{S_0}{4\rho} -\frac{(4r + \sigma^2)}{8\rho}\left(\frac{S_0}{2} + \sqrt{sS_0}\right)t\\
            u^{(2)}(s,t) &= s - \frac{\sqrt{s S_0}}{\rho} - \frac{S_0}{4\rho} -\frac{(4r + \sigma^2)}{8\rho}\left(\frac{S_0}{2} + \sqrt{s S_0}\right)t\\
            &- \frac{(4r + \sigma^2)^2}{64\rho}\left(S_0 + \sqrt{s S_0}\right)\frac{t^2}{2!} - \frac{(4r + \sigma^2)^2}{512\rho}\left(\sigma^2 S_0\right)\frac{t^3}{3!}\\
            &\vdots\\
        \end{split}
    \end{equation}
which converges slowly to the exact solution \eqref{eq:black-scholes_exact}.

Next, the GVIM produces the following iterates
   \begin{equation}
        \label{eq:black_scholes_gvim_approximants}
        \begin{split}
            u^{(0)}(s,t) &= s - \frac{\sqrt{s S_0}}{\rho} - \frac{S_0}{4\rho}\\
            u^{(1)}(s,t) &= s - \frac{1}{8r\rho}\left(\frac{S_0}{2} + \sqrt{sS_0}\right) \left(\me^{-rt} (4r+\sigma^2) - \sigma^2\right) - \frac{\sqrt{s S_0}}{2 \rho}\\
            u^{(2)}(s,t) &= s - \frac{1}{4\rho} \left(1-\frac{\sigma^2}{2r} + \frac{\sigma^4}{16r^2}\right)\left(\sqrt{s S_0} - \frac{S_0 \sigma^2}{16r}\right) -\frac{S_0(4r+\sigma^3)^2}{1024  r^3\rho}\me^{2rt}\\
            &- \frac{4r+\sigma^2}{64r^2\rho}\left((4r-\sigma^2) (S_0 + \sqrt{s S_0}) + 8r \sqrt{sS_0}\right) \me^{rt}\\
            &+ \frac{16r^2 - \sigma^4}{512r^2\rho}\left(8r\sqrt{sS_0} - \sigma^2 S_0\right) t \me^{rt}\\
            &\vdots\\
        \end{split}
    \end{equation}
    
\begin{figure}[!ht]
    \centering
    \begin{subfigure}{0.48\textwidth}
        \includegraphics[width=\linewidth]{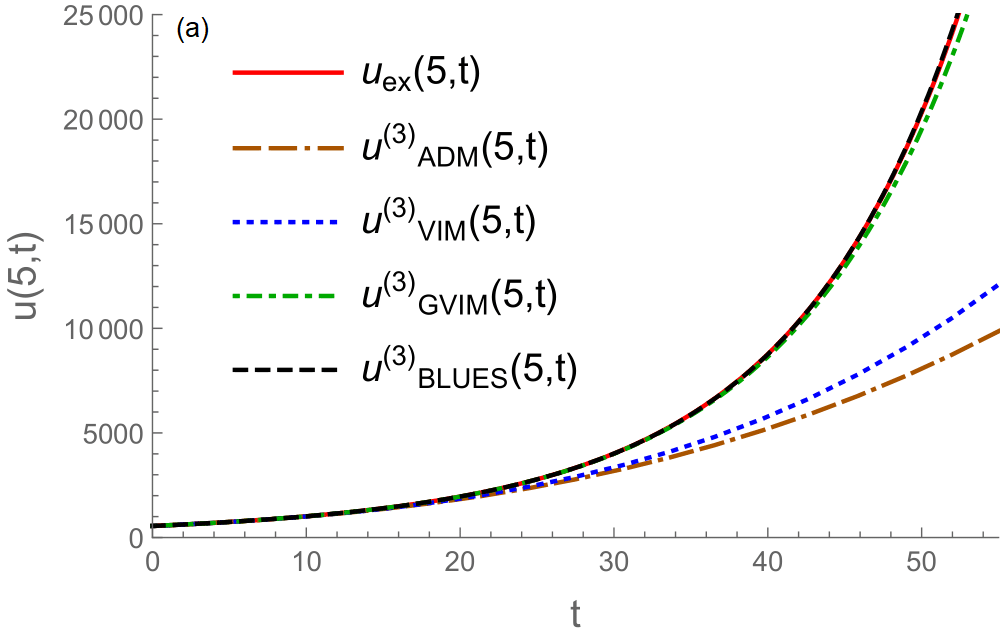}
    \end{subfigure}
    \begin{subfigure}{0.48\textwidth}
        \includegraphics[width=\linewidth]{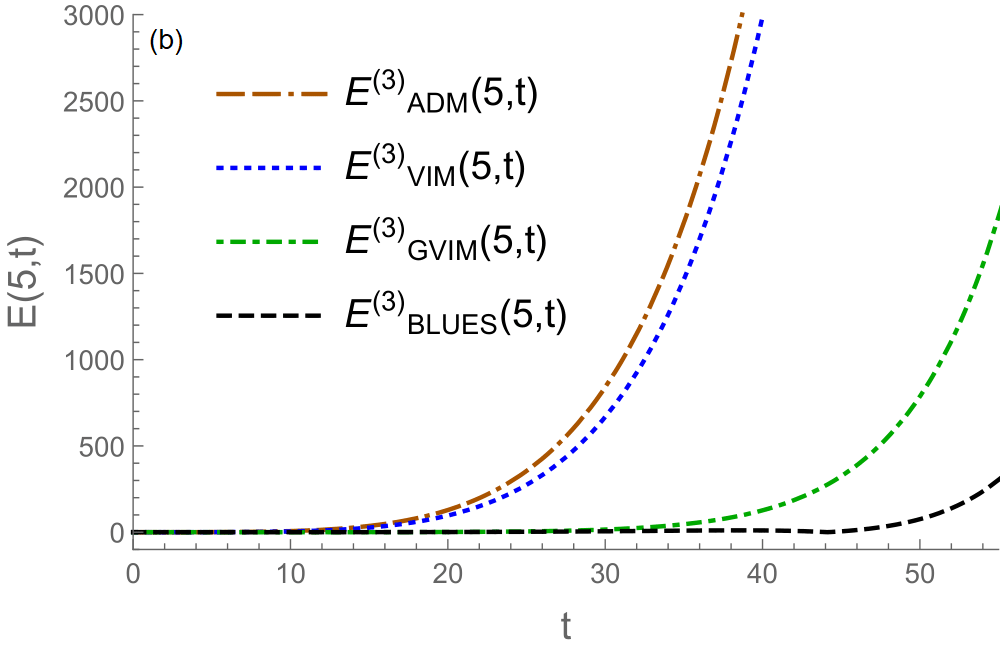}
    \end{subfigure}
    \caption{Black-Scholes equation. \textbf{(a)} Exact solution $u_{ex}$ (red, full line) and approximants $u^{(3)}$ in third iteration or third order. \textbf{(b)} Difference $E^{(3)}$ between the exact solution \eqref{eq:black-scholes_exact} and the approximant of order $n =3$ for the different methods: ADM \eqref{eq:black_scholes_ADM_approximants} (orange, dot-dash-dashed line), VIM \eqref{eq:black_scholes_vim_approximants} (blue, dotted line), GVIM \eqref{eq:black_scholes_gvim_approximants} (green, dot-dashed line), BLUES \eqref{eq:black_scholes_blues_approximants} (black, dashed line). The asset price coordinate is fixed at $s = 5$. Reduced (dimensionless) values of the parameters are $r=0.06$, $\sigma = 0.4$, $\rho = -0.01$ and $S_0 = 4$.}
    \label{fig:black_scholes_comparison_error}
\end{figure}

Finally, we study the BLUES method. As usual, we first rewrite equation \eqref{eq:pde_black_scholes} with the inclusion of a source $\psi(s,t) = f(s)\delta(t)$, i.e.,
\begin{equation}
    \label{eq:pde_black_scholes_rewrite}
    u_t + \frac{\sigma^2 s^2}{2} u_{ss} \left(1+2\rho \,s \,u_{ss}\right) + r s\, u_s - r u = \psi\, ,
\end{equation}
and consider the associated linear operator we have used in the previous examples together with the source $\psi(s,t)$,   
\begin{equation}
    \label{eq:black_scholes_blues_linear_operator}
    \lopt u = u_t - r u = \psi
\end{equation}
with Green function,
\begin{equation}
    \label{eq:black_scholes_blues_greenfunction}
    G(t) = \Theta (t) \me^{rt}\, .
\end{equation}    
Note that in this example, the linear operator is chosen judiciously by not only dropping the nonlinear term but some linear terms as well. Hence, the residual, whose action is defined through
\begin{equation}
    \label{eq:black_scholes_blues_residual}
    \mathcal{R}_{s}\, u = - \frac{\sigma^2 s^2}{2}u_{ss}\left(1+2 \rho \,s \,u_{ss}\right) -r s\, u_s,
\end{equation}
still contains two linear terms. The zeroth approximant is the convolution of the BLUES function \eqref{eq:black_scholes_blues_greenfunction} and the source $\psi(s,t)$, 
\begin{equation}
    \label{eq:black_scholes_blues_zerothorder}
    \begin{split}
    u^{(0)}(s,t) &= \int\limits_{0^-}^t \mathrm{d}t'\, G(t- t')\psi(s,t') = \left(s - \sqrt{s S_0}/\rho - S_0/(4\rho)\right) \me^{rt} \, .
    \end{split}
\end{equation}    
The BLUES function method  generates the following sequence of approximants
\begin{equation}
    \label{eq:black_scholes_blues_approximants}
    \begin{split}
         u^{(0)}(s,t) &= \left(s - \frac{\sqrt{s S_0}}{\rho} - \frac{S_0}{4\rho}\right) \me^{rt}\\
         u^{(1)}(s,t) &= \left(s - \frac{\sqrt{sS_0}}{\rho} -\frac{S_0(4r+\sigma^2)}{16r\rho}\right)\me^{rt} - \left(rs - \frac{(4r-\sigma^2)}{8\rho}\sqrt{sS_0}\right) \me^{rt}t\\
         &- \frac{\sigma^2 S_0}{16r\rho}\me^{2rt}\\
         u^{(2)}(s,t) &=\left(s - \frac{\sqrt{sS_0}}{\rho} - \frac{S_0 \left(4r-\sigma^2\right)^2 (8r-\sigma^2)}{512r\rho^3}\right)\me^{rt} - \left(rs -\frac{(4r-\sigma^2)}{8\rho}\sqrt{sS_0}\right) \me^{rt}t\\
         &+\left(r^2 s - \frac{(4r-\sigma^2)^2}{64\rho}\sqrt{sS_0}\right)\frac{\me^{rt}t^2}{2} - S_0\left(\frac{80r^2 \sigma^2 - 16 r \sigma^4 + \sigma^6}{512r^3\rho}\right)\me^{2rt}\\
         &+ S_0\left(\frac{48r^2\sigma^2 - 16r\sigma^4 +\sigma^6}{512r^2\rho}\right)\me^{2rt}t - S_0\left(\frac{\sigma^2 (4r-\sigma^2)^2}{512r\rho}\right)\frac{\me^{2rt}t^2}{2}\\
          &\vdots\\
    \end{split}
\end{equation}
In Fig. \ref{fig:black_scholes_comparison_error} we compare the results from each of the above methods and also compare their errors, at the level of the 3rd approximant or 3rd order ($n=3$).  
Note that we have not chosen an explicit value for the expiration time $T$ and considered $t\in[0,\infty)$, i.e., $T\rightarrow\infty$. If one were to fix $T>0$ at a finite value, it is obvious that the accuracy of the approximate solutions for all of the above procedures decreases for $t\rightarrow T$, i.e., for increasing remaining time until end of contract.
    
\section{Diffusion equation with general nonlinearity}
\label{sec:general}

We now set the stage for the analysis of a nonlinear PDE associated with a simple physical model for the growth of an interface between two fluids that are subject to shear flow, by first considering a more general nonlinear PDE from a technical viewpoint. The heat equation with diffusion constant $D>0$ and general nonlinearity $u^m u_x^n$, where $m,n \geq0$ is given by the PDE,
\begin{equation}
    \label{eq:heat}
    \nlop u = u_t - D u_{xx} - u^m u_x^n =0,
\end{equation}
with Gaussian initial condition $u(x,0) = f(x)$
\begin{equation}
    \label{eq:initial_condition_gaussian}
   f(x) = \frac{\mathrm{e}^{-x^2/2\sigma^2}}{\sqrt{2\pi\sigma^2}}
\end{equation}
and boundary conditions $u(|x|\rightarrow \infty,t)=0$. As before, we adopt the notation $\nlop u$ to denote the nonlinear operator acting on  $u(x,t)$. The associated linear PDE of our choice is the one-dimensional heat equation describing normal diffusion,
\begin{equation}
    \label{eq:diffusion}
    \lop u = u_t - D u_{xx} = 0,
\end{equation}
with the same initial condition and the same boundary conditions. This linear PDE has Green function 
\begin{equation}
    \label{eq:green}
    G(x,t) = \frac{\mathrm{e}^{-\frac{x^2}{4Dt}}}{\sqrt{4\pi Dt}}.
\end{equation}
In the small time limit $t\rightarrow0$, the Green function \eqref{eq:green} approaches a Dirac-delta distribution $\delta(x)$. The solution to the diffusion equation with the Gaussian initial condition $f(x)$  can be calculated by convoluting $f(x)\delta(t)$ with the kernel $G(x,t)$,
\begin{equation}
    u^{(0)}(x,t) = \int\limits_{0^-}^t\int_\mathbb{R}\mathrm{d}y\,\mathrm{d}s\, G(x-y,t-s)f(y)\delta(s).
\end{equation}
Integrating over time and space gives
\begin{equation}
\label{eq:convolution}
\begin{split}
    u^{(0)}(x,t) = \int_\mathbb{R}\mathrm{d}y \,G(x-y,t)f(y)  
    = \frac{\mathrm{e}^{-x^2/2\Sigma^2(t)}}{\sqrt{2\pi \Sigma^2(t)}}\, ,
\end{split}
\end{equation}
which is itself a decaying Gaussian with mean zero and with variance $\Sigma^2(t) \equiv \sigma^2 + 2Dt$. This solution $u^{(0)}$ serves as the zeroth iteration in the BLUES scheme. One now considers the residual operator $\rop = \lop  - \nlop $ which can be applied to the zeroth approximant \eqref{eq:convolution},
\begin{equation}
    \label{eq:residual_general}
    \begin{split}
    \rop  u^{(0)}(x,t) = \left(u^{(0)}\right)^m\left(u_x^{(0)}\right)^n 
        = (-1)^{n} \frac{x^n \mathrm{e}^{-(m+n)x^2/2\Sigma^2(t)}}{(2\pi)^{\frac{m+n}{2}}\Sigma(t)^{m+3n}}
    \end{split}
\end{equation}
Convoluting the previous expression with the Green function \eqref{eq:green} results in 
\begin{equation}
    \begin{split}
    u^{(1)} (x,t) - u^{(0)} (x,t)= \frac{(-1)^{n}}{(2\pi)^{\frac{m+n+1}{2}}}\int\limits_{0^-}^t\mathrm{d}s\frac{\mathrm{e}^{-x^2/2S^2(t,s)}}{\sqrt{2D(t-s)}  \Sigma(s)^{m+3n}}\int_\mathbb{R}\mathrm{d}y\, y^n \mathrm{e}^{-\alpha(t,s)\,(y-c(t,s)\,x)^2}\, ,
    \end{split}
\end{equation}
where $S^2(t,s) \equiv 2D(t-s) + \Sigma^2(s)/(m+n)$, which can be interpreted as a variance. Further, $c(t,s) \equiv (\Sigma^2(s)/S^2(t,s))/(m+n)$ and $\alpha (t,s)\equiv (m+n) (S^2(t,s)/\Sigma^2(s)) /(4D(t-s))$. The spatial integral can be calculated exactly

\begin{equation}
\label{eq:lambda}
\begin{split}
    \Xi(x,t,s,m,n) &\equiv \int_\mathbb{R}\mathrm{d}y\,y^n \mathrm{e}^{-\alpha(t,s)\,(y-c(t,s)\,x)^2}\\
    &= 
    \alpha^{-\frac{n+1}{2}}\begin{cases}
        \Gamma\left(\frac{n+1}{2}\right){}_1F_1\left(-\frac{n}{2}, \frac{1}{2},-\alpha c^2x^2\right), &  \text{$n$ even}\\
        n\sqrt{\alpha c^2 x^2}\,\Gamma\left(\frac{n}{2}\right){}_1F_1\left(-\frac{n-1}{2},\frac{3}{2},-\alpha c^2x^2\right), &  \text{$n$ odd}\, ,
    \end{cases}
\end{split}
\end{equation}
where $\Gamma(n)$ is the gamma function and $_1F_1(a,b,z)$ is the confluent hypergeometric function of the first kind \cite{abramowitz+stegun}. This spatial integral can equivalently be expressed in terms of the Hermite polynomials $H_n(z)$ in the following way, 
\begin{equation}
    \Xi(x,t,s,m,n) \equiv \left(\frac{-i}{2}\right)^n \sqrt{\frac{\pi}{\alpha(t,s)^{n+1}}}H_n\left(i \sqrt{\alpha(t,s)}c(t,s)x\right)\,.
\end{equation}
We list here the following useful properties for the hypergeometric functions and for the Hermite polynomials:
\begin{align}
  _1F_1(0,b,z) &= 1 \label{eq:hypergeo1}\\
  _1F_1(-1,b,z) &= 1-\frac{z}{b} \label{eq:hypergeo2}\\
  H_1(z) &= 2z \label{eq:hermite1}\\
  H_2(z) &= 4z^2 -2\,. \label{eq:hermite2}
\end{align} 

\noindent The first correction to the zeroth approximant \eqref{eq:convolution} now becomes
\begin{equation}
\label{eq:general_first}
    u^{(1)}(x,t) - u^{(0)}(x,t) = 
    \frac{(-1)^{n}}{(2\pi)^{\frac{m+n+1}{2}}}\int\limits_{0^-}^t\mathrm{d}s\,\frac{\mathrm{e}^{-x^2/2S^2(t,s)}}{\sqrt{2D(t-s)}  \Sigma(s)^{m+3n}}\Xi(x,t,s,m,n)
\end{equation}

\noindent For some choices of $(m,n)$ this can be simplified greatly. In the next section we discuss a physical system which features two such cases combined, $(1,1)$ and $(0,2)$.

\section{Interface growth under shear}
\label{sec:interface}

We propose a minimalistic model for the growth of an interface between two fluids near two-phase coexistence and subject to an externally imposed shear flow. On the one hand, we exploit the finding that the growing interface between a stable and an unstable domain in a kinetic Ising model at low temperature can be described by including in the effective growth equation a Kardar-Parisi-Zhang (KPZ) nonlinearity which allows for lateral growth \cite{KPZ,Devillard_1992,Krug1991,barabasi_stanley_1995}. On the other hand, we make use of the growth equation proposed for studying interface fluctuations under shear flow, including a Burgers type of nonlinearity \cite{burgers} which allows for a background linear shear flow imposed on the phase-separated fluid \cite{bray2001,bray2001_2}. We combine the two growth equations but limit ourselves to the minimal setting of two-dimensional systems (i.e., a one-dimensional interface) and the deterministic version of the equation. We ignore thermal noise and postpone an application to the stochastic DE until later work.

Our starting point is, as usual, the Edwards-Wilkinson equation for interface growth \cite{Edwards1982}, which, in its deterministic version, reads
\begin{equation}
    \label{eq:EW}
    h_t - D h_{xx}   = 0\, ,
\end{equation}
where $h(x,t)$ is the height of an interface that fluctuates, measured relative to a (horizontal) straight reference line (along $x$). This reference line is co-moving with the growing interface and therefore a velocity term $v$ is omitted in \eqref{eq:EW}. $D$ is a diffusion coefficient (proportional to the interfacial tension whose action is to smoothen the interface).

\begin{figure}[htp]
    \centering
    \includegraphics[width=0.8\linewidth]{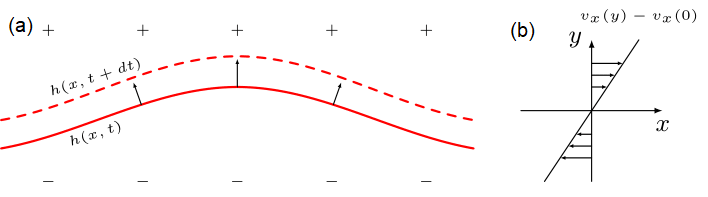}
    \caption{Cartoon of a coarse-grained growing interface, a density contour of which is described by a collective coordinate $h(x,t)$, between (stable) ``$-$" and (unstable) ``$+$" domains in the 2d Ising model representation of a phase-separated fluid. \textbf{(a)} In the absence of flow the interface advances mainly in the direction normal to its tangent. \textbf{(b)} The fluid as a whole is subject to an externally imposed shear flow with linear profile $v_x(y)$.}
    \label{fig:interfacecartoon}
\end{figure}

A cartoon of the physical setting is shown in Fig. \ref{fig:interfacecartoon}. Following Bray {\em et al.} \cite{bray2001,bray2001_2} we include an externally imposed shear flow. The motivation, in part, for this was that there is an interesting subtle competition between the smoothing of an interface under shear and the roughnening of an interface under thermal noise. Later studies elucidated interface confinement under shear using Monte Carlo simulation \cite{Smith2008,Smith_2008_2}. Incorporating a (horizontal) shear velocity profile $v_x (y)$ amounts to invoking the total time derivative,
\begin{equation}
    \label{eq:totalderivative}
    h_t \rightarrow \frac{dh}{dt} = h_t + v_x(h) h_x,
\end{equation}
since $h$ is the $y$-coordinate of the interface position. For shear flow, $v_x(h) $ is a linear function $Ah+B$ and we can choose  a reference frame co-moving at the mean velocity, so $B=0$. We thus add a Burgers convective nonlinearity to the PDE.

Next, following Devillard and Spohn \cite{Devillard_1992} we recognize that the interface growth, ignoring the lattice anisotropies of the model, is in the direction normal to the local tangent. This growth, in which a stable domain overtakes an unstable one, is driven by a pressure difference, or chemical potential difference, with respect to two-phase coexistence (i.e., a non-zero external magnetic field in the Ising model). Incorporating this lateral growth amounts to invoking the KPZ geometric correction,
\begin{equation}
    \label{eq:KPZgeometry}
    v  \rightarrow v + \frac{v}{2} (h_x)^2,
\end{equation}
where $v$ is the velocity of the growing interface. Since the term $v$ is already absorbed in \eqref{eq:EW} we need to add only the gradient-squared term to the PDE. Altogether we obtain the nonlinear PDE
 \begin{equation}
    \label{eq:interfacegrowthBKPZ}
    h_t + A h\, h_x = D h_{xx}  + \frac{v}{2} h_x^2 \, ,
\end{equation}
where $A$ is the shear rate.

This PDE combines the Burgers and KPZ nonlinearities but, we recall, ignores thermal noise. When taken separately, each of these two nonlinearities amount to exactly solvable PDEs, but to our knowledge not when combined. This makes it worthwhile to derive a useful analytical approximant to the solution of the combined equation. Note that in our physical context extra terms proportional to $h$ or $h^2$ are not present in \eqref{eq:interfacegrowthBKPZ} because in the absence of shear flow we require translational invariance of the growth equation along the $y$-direction. In addition, we require translational invariance along $x$. Also note that in terms of the scaling properties of interface growth the Burgers term is the dominant perturbation \cite{bray2001,bray2001_2} and the KPZ term is subsidiary. We do not discuss these properties here.

There is an alternative route to the  PDE \eqref{eq:interfacegrowthBKPZ} which is worth pointing out. One may start from the stochastic KPZ equation for interface growth and couple it to the stochastic Navier-Stokes (NS) equation for the velocity field $v$, by replacing the time derivative in KPZ by the total time derivative, as in \eqref{eq:totalderivative}, and invoking the NS equation for  $v$. This system of coupled DEs was proposed and studied in \cite{Antonov_2020}. If, in that system, one ignores the random force in the stochastic NS equation and imposes a (deterministic) shear flow velocity profile, and if one also ignores thermal noise in the KPZ equation, one arrives again at \eqref{eq:interfacegrowthBKPZ}.

We now proceed to the calculations and adapt the notation slightly in order to be conform with that of previous sections. We define the nonlinear operator, acting on the function $u(x,t)$,
 \begin{equation}
    \label{eq:BKPZ}
  \nlop u =   u_t - D u_{xx} + \alpha u\, u_x + \beta u_x^2 \, ,
\end{equation}
with $\alpha$ and $\beta$ real parameters.
For the linear operator $\lop$ we choose the entire linear part of $\nlop$, which is the linear diffusion operator. The residual operator $\rop$ (cf.  Section \ref{sec:general}), is then defined through
\begin{equation}
    \label{eq:interface_residual}
    \rop u = -\alpha u\, u_x -\beta u_x^2
\end{equation}
By doing so, the nonlinear problem would be suited to be tackled by {\em perturbation theory} (PT), if the terms that feature the parameters $\alpha$ and $\beta$ can be considered to be small compared to the terms of the linear part. This brings us in position to compare the BLUES iteration, which is non-perturbative, to a direct perturbation expansion, keeping in mind that the former makes no assumptions on the magnitude of the nonlinear terms. What we find is akin to our observations in the treatment of ODEs \cite{Berx_2020}. The BLUES iteration generates a sequence that is in general different from summing up the terms a series expansion, except possibly in the first iteration in which the BLUES result may coincide with that of 1st-order PT. 

We consider two different initial conditions, corresponding to distinct physical situations. The first is a single (Gaussian) interface protrusion or ``bump", for which we will illustrate the method at the level of the zeroth and first iteration only, and show its close similarity to 1st-order PT. The second one is a (sinusoidal) periodic interface front, for which we will study the time evolution to higher level in the iteration scheme.  For that case, we will perform a detailed comparison of the results from ADM, VIM, GVIM, BLUES and PT. 

\subsection{Gaussian initial condition}
First, we will consider the situation of a solitary interface bump that can be modeled by a Gaussian initial condition $u(x,0) = f(x)$,  given in equation \eqref{eq:initial_condition_gaussian}. We assume the boundary conditions $u(|x|\rightarrow\infty) = 0$.  The associated linear PDE is the heat equation \eqref{eq:diffusion}. The zeroth approximant is now the decaying Gaussian  solution \eqref{eq:convolution} of the linear equation. Using equation \eqref{eq:general_first} twice, once for the convective nonlinearity (Burgers) and once for the nonlinear lateral growth (KPZ), the first approximant can be calculated analytically. We report here the result (a detailed calculation can be found in Appendix \ref{app:interfaces}),
\begin{equation}
    \label{eq:interface_first_solved}
    \begin{split}
         u^{(1)}(x,t) &= \frac{\mathrm{e}^{-x^2/2\Sigma^2(t)}}{\sqrt{2\pi \Sigma^2(t)}} + \frac{\beta}{4\pi D}\left[\frac{ \mathrm{e}^{-x^2/\Sigma^2(t)}}{\Sigma^2(t)}-\frac{ \mathrm{e}^{-x^2/\Sigma^2(2t)}}{\Sigma(2t)\sigma}\right] \\
         &+\frac{\alpha}{4D\sqrt{2\pi}}\left[\frac{\mathrm{e}^{-x^2/2\Sigma^2(t)}}{\Sigma(t)}\left(\erf{\left(\frac{x}{\sqrt{2}\Sigma(t)}\right)} -\erf{\left(\frac{\sigma x}{\sqrt{2}\Sigma(t)\Sigma(2 t)}\right)}\right)\right]
    \end{split}
\end{equation}
Note that the effects introduced by the convective nonlinearity contain only odd functions of $x$, and the effects introduced by the nonlinear growth contain only even functions of $x$. In the first iteration the effect of nonlinearity is a simple superposition of the individual nonlinear effects, i.e., nonlinear convection and nonlinear growth. Only in higher iterations does the interplay (mixing) between these different effects take place. 

At this level of approximation, the BLUES approximant $u^{(1)}$ coincides with the result of straightforward PT to first order in $\alpha$ and $\beta$. This is not surprising in view of the fact that the chosen residual operator coincides with the nonlinear part of the differential operator, which is precisely the ``perturbation" when $\alpha$ and $\beta$ are considered small. We have also performed the ADM and VIM calculations for this case. These methods are, however, not suitable here because they produce large oscillations that grow uncontrollably both in time and in higher orders of approximation. We will return to these methods when we consider a periodic interface undulation.

In the first iteration of the nonlinear problem we obtain,
\begin{equation}
    \label{eq:bump_excess_gaussian}
    \begin{split}
        \int_\mathbb{R} dx\, u^{(1)}(x,t) &=
        1+\frac{\beta}{4D\sqrt{\pi}}\left(\Sigma^{-1}(t) -\sigma^{-1}\right)\\
    \end{split}
\end{equation}
This is a non-decreasing function of time for $\beta<0$, hence the bump grows as a consequence of the lateral growth correction, even when there is no overall (vertical) growth along $y$ in the co-moving frame.  Note that the parameter $\alpha$ does not enter the equation. The shear flow only moves particles along $x$ and does not influence the bump size but only its shape. 

\begin{figure}[htp]
    \centering
    \includegraphics[width=0.8\linewidth]{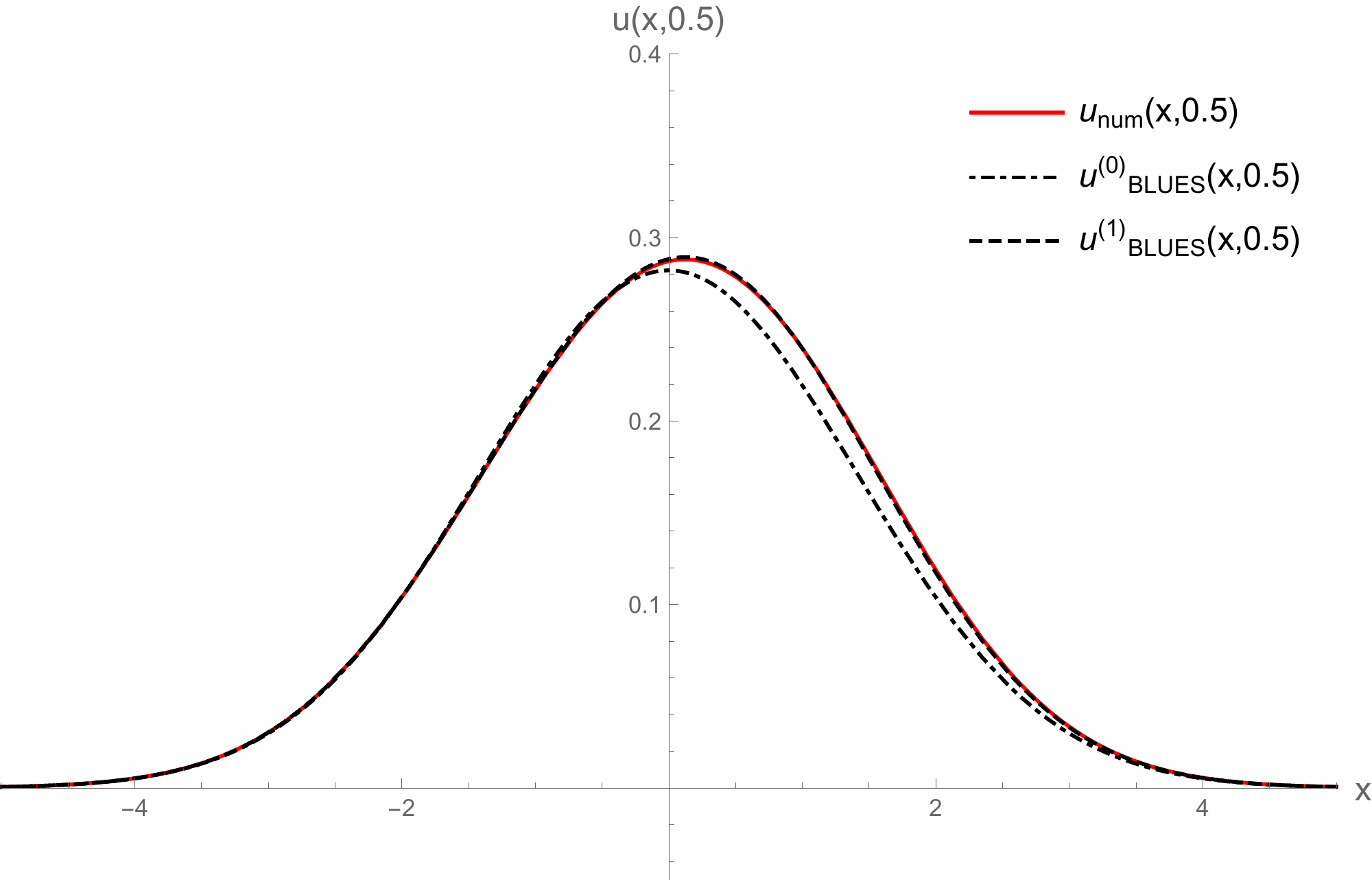}
    \caption{Solitary interface bump  at time $t=1/2$. The numerical solution (red line) is compared with the zeroth (dot-dashed line) and first (dashed line) BLUES  approximants \eqref{eq:interface_first_solved}. Reduced (dimensionless) values of the parameters are $D=\sigma=\alpha=1$ and $\beta=-1$. These results coincide with, respectively, those of standard zeroth and 1st-order PT in the parameters $ \alpha$ and $\beta$. In the course of time the bump distorts to the right (for $\alpha >0$) and grows somewhat (for $\beta < 0$) until its size saturates.}
    \label{fig:Interface_Gaussian_space}
\end{figure}

\begin{figure}[htp]
    \centering
    \includegraphics[width=0.8\linewidth]{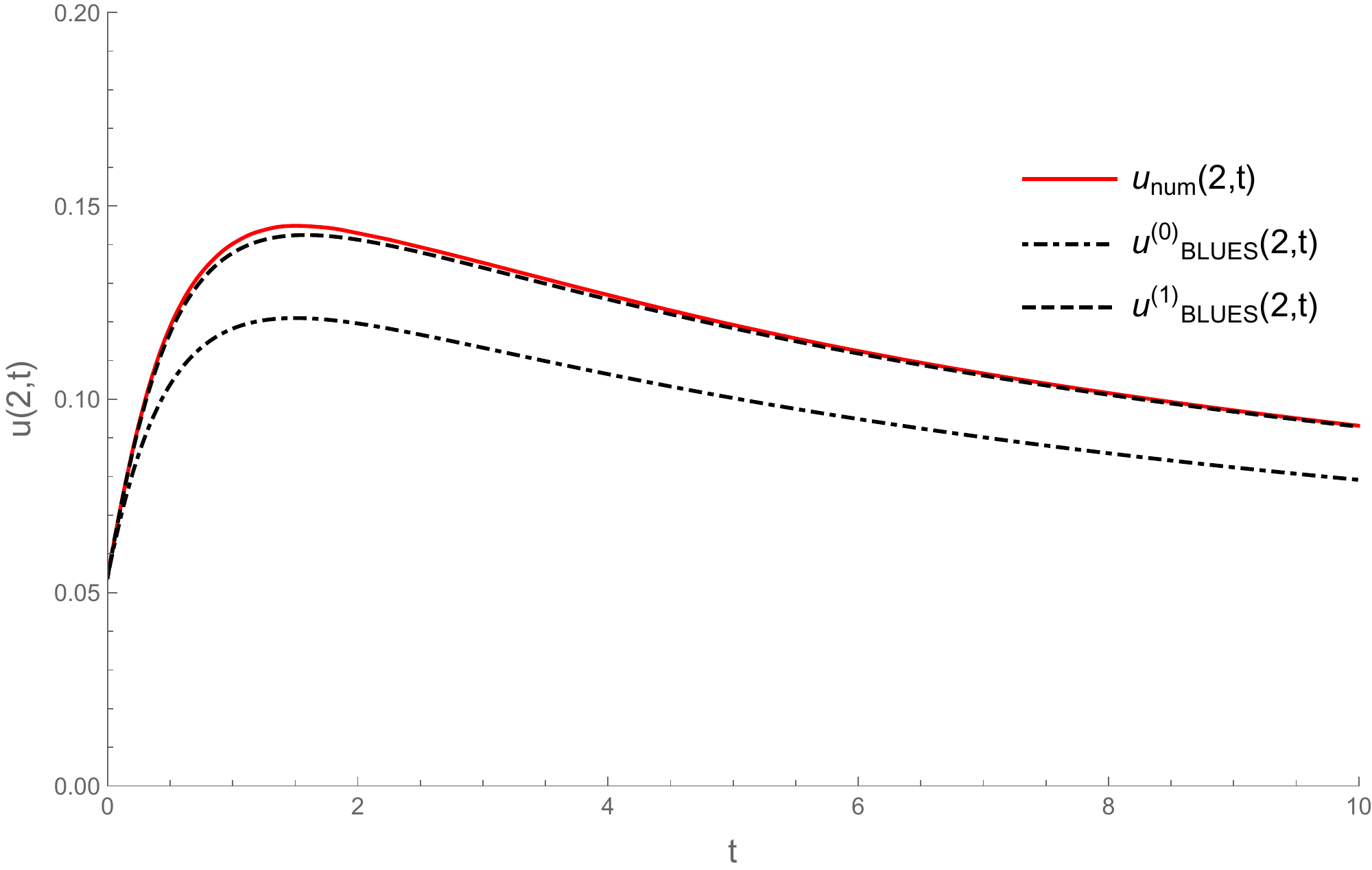}
    \caption{Solitary interface bump time evolution at position $x=2$. The numerical solution for  $u$ (red line) is compared with the zeroth (dot-dashed line) and first (dashed line) BLUES  approximants \eqref{eq:interface_first_solved}. These results coincide with, respectively, those of standard zeroth and 1st-order PT in the parameters $ \alpha$ and $\beta$. Reduced (dimensionless) values of the parameters are $D=\sigma=\alpha=1$ and $\beta=-1$.}
    \label{fig:Interface_Gaussian_time}
\end{figure}

In Fig. \ref{fig:Interface_Gaussian_space} the short-time shift of the bump is illustrated (snapshot at $t=1/2$), as obtained with zeroth and 1st iteration BLUES as well as zeroth and 1st-order PT, which gives the same results. In Fig. \ref{fig:Interface_Gaussian_time} the time evolution at fixed position ($x=2$) is shown, using the zeroth and first BLUES approximants. In both figures the results are compared with the numerically exact solution.

\subsection{Space-periodic initial condition}
\begin{figure}[htp]
    \centering
    \includegraphics[width=0.75\linewidth]{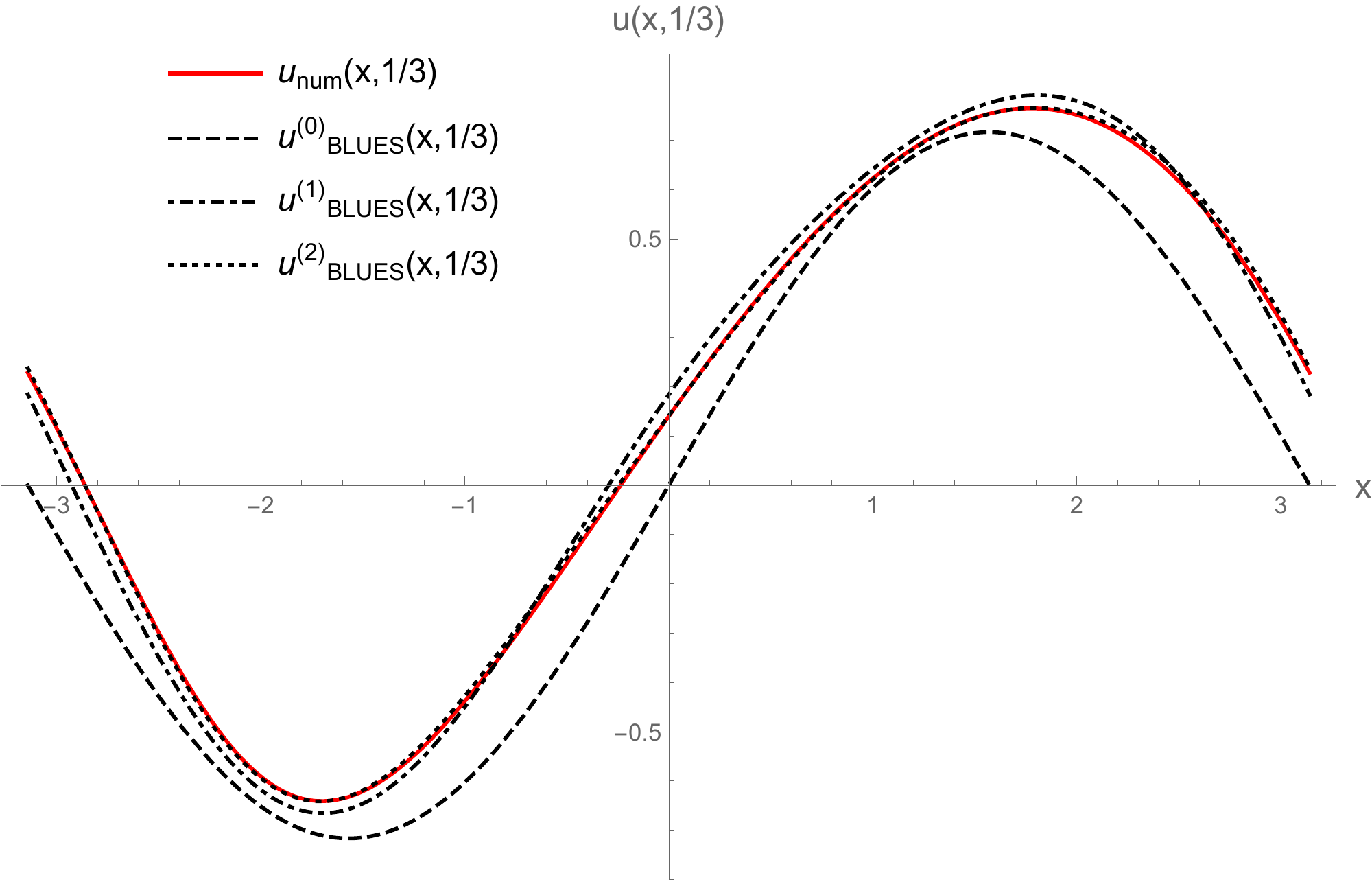}
    \caption{Periodic interface contour at time $t=1/3$. The numerical solution (red line) is compared with the $n=0,1$ and $2$ BLUES  approximants. The second approximant nearly coincides with the numerical solution at this resolution. Parameter values are $D=\alpha=1$ and $\beta=-1$.}
    \label{fig:interface_gaussian_space_orders}
\end{figure}

\begin{figure}[htp]
    \centering
    \includegraphics[width=0.75\linewidth]{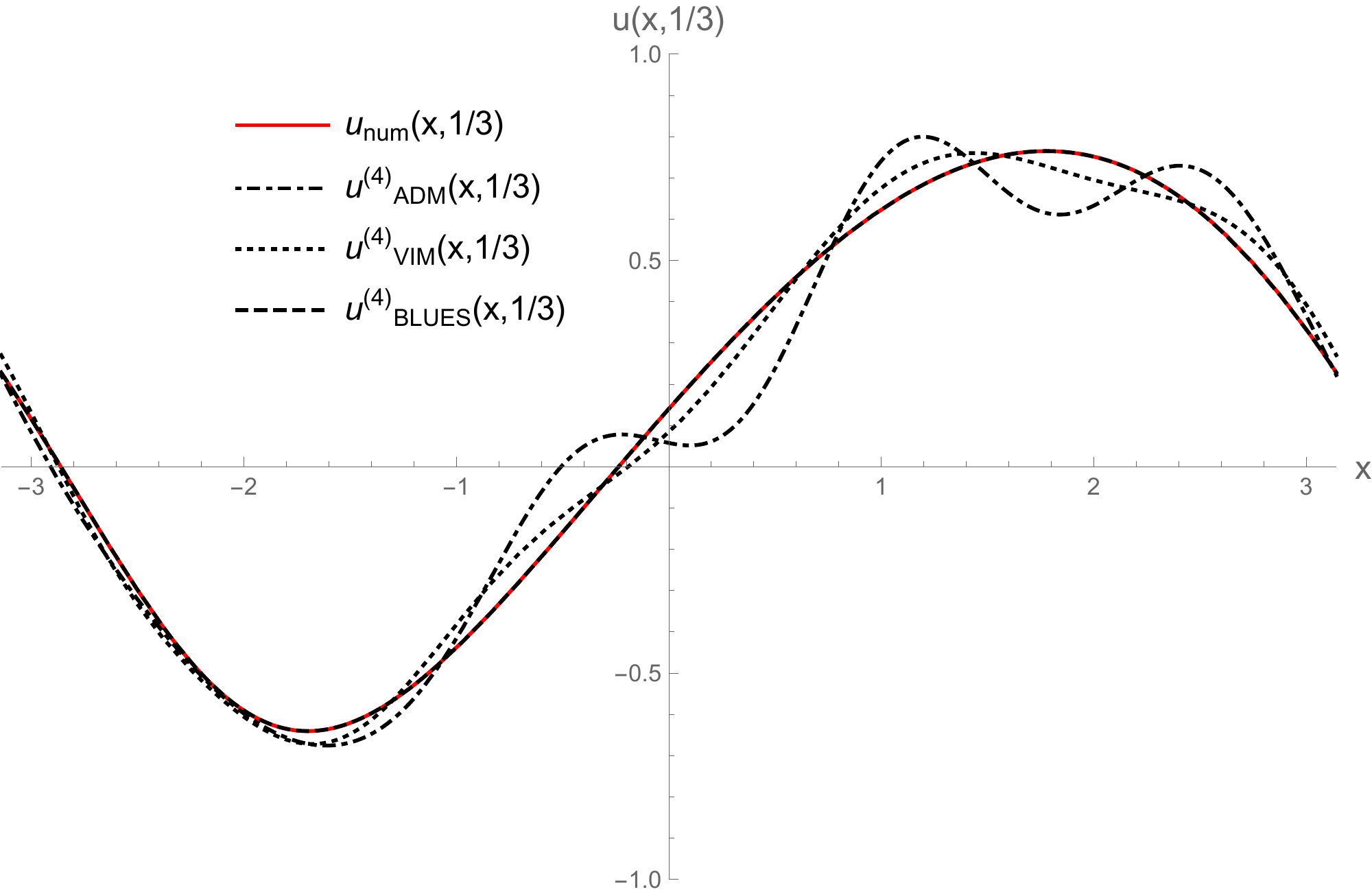}
    \caption{Periodic interface contour at time $t=1/3$. The numerical solution (red line) is compared with the $n=4$ BLUES method approximant (dashed line) and the $n=4$ approximant of the VIM and the ADM (respectively dot-dashed and dotted lines). At this resolution the fourth BLUES approximant falls on top of the numerical solution. Parameters are $D=\alpha=1$ and $\beta=-1$.}
    \label{fig:interface_gaussian_space}
\end{figure}

For convenience and simplicity, in this example we will work with dimensionless variables $x$ and $t$, as well as dimensionless $u$, $D$, $\alpha$ and $\beta$. To study a space-periodic interface contour, we can choose the following trigonometric initial condition $f(x)$
\begin{equation}
    \label{eq:interface_sin_ic}
    f(x) = \sin  x
\end{equation}
and examine the behavior of solutions of the suitably rescaled version of equation \eqref{eq:BKPZ} on the real line. The zeroth approximant is the convolution integral of the Green function \eqref{eq:green} with \eqref{eq:interface_sin_ic}, 
\begin{equation}
    \label{eq:interface_sin_convolution}
    u^{(0)}(x,t) = \me^{-Dt}\sin x\, .
\end{equation}
One can now apply the residual operator \eqref{eq:interface_residual} to \eqref{eq:interface_sin_convolution}. After simplifying the result by using trigonometric power reduction identities, the residual is
\begin{equation}
    \label{eq:interface_sin_residual_convolution}
    \rop u^{(0)}(x,t) = -\frac{\me^{-2 Dt}}{2}\left(\alpha \sin 2x +\beta \cos 2x+\beta\right)
\end{equation}
The first  approximant to the solution of equation \eqref{eq:BKPZ} can be calculated by convoluting the residual \eqref{eq:interface_sin_residual_convolution} with the Gaussian Green function, making use of the following identities
\begin{equation}
    \label{eq:trigonometric_integral_identities}
    \begin{split}
        \int\limits_\mathbb{R} dy\, \frac{\mathrm{e}^{-\frac{(x-y)^2}{4D(t-s)}}}{\sqrt{4\pi D(t-s)}} \sin{ay} &= \me^{-a^2 D(t-s)}\sin{ax}\\
        \int\limits_\mathbb{R} dy\, \frac{\mathrm{e}^{-\frac{(x-y)^2}{4D(t-s)}}}{\sqrt{4\pi D(t-s)}} \cos{ay} &= \me^{-a^2 D(t-s)}\cos{ax}
    \end{split}
\end{equation}
Hence, the first approximant is 
\begin{equation}
    \label{eq:interface_sin_first}
    u^{(1)}(x,t) = \me^{-Dt}\sin x +\frac{\me^{-2Dt}\left(\me^{-2 Dt}-1\right) }{4 D}\left[\alpha\sin 2x + \beta\cos 2x +\beta\me^{2Dt}\right]
\end{equation}
Higher approximants can be calculated with moderate effort. In Fig. \ref{fig:interface_gaussian_space_orders} we show the first three BLUES approximants together with the numerically exact solution for a fixed time $t=1/3$. Next, in Fig.\ref{fig:interface_gaussian_space} we compare the numerical solution and the fourth BLUES approximant with the 4th-order VIM and ADM results at $t=1/3$.

Let us now juxtapose  BLUES approximant of the second iteration with a 2nd-order solution obtained from PT. The first(-order) approximants of both methods coincide exactly so we will consider the following perturbation expansion $u_P$ for the solution 
\begin{equation}
    \label{eq:perturbative_solution_definition}
    \begin{split}
        u_{PT}(x,t) &= u_{0,0}(x,t) + \alpha u_{1,0}(x,t) + \alpha^2 u_{2,0}(x,t)+ \beta u_{0,1}(x,t)\\
        & + \beta^2 u_{0,2}(x,t) + \alpha\beta u_{1,1}(x,t) + \mathcal{O}(\alpha^m\beta^n), \;\;m+n =3,
    \end{split}
\end{equation}
and we assume, within PT, to avoid ambiguity, that $\alpha$ and $\beta$ are of the same order of magnitude.
Performing the expansion and solving the resulting linear PDEs yields the expressions given in Appendix \ref{app:fourier} for the  perturbative solution $u^{(2)}_{PT}$  up to, and including, second order in $\alpha$ and $\beta$.

Note that PT generates terms of second order in $\alpha$ and $\beta$, i.e., $\alpha^2$, $\beta^2$ and $\alpha\beta$, and Fourier modes up to and including the third harmonic (with respect to the period of the initial condition). In contrast, in the second iteration the BLUES function method does not yet provide the exact coefficients of the 2nd-order terms. Furthermore, this method also generates terms of higher order in $\alpha$ and $\beta$, e.g., $\alpha^3$, $\beta^3$, $\alpha^2\beta$, etc., and Fourier modes of the fourth harmonic are also already present in the second approximant. We provide also the full expressions of the 2nd approximant in Appendix \ref{app:fourier} and compare them quantitatively with PT. 

\begin{figure}[htp]
    \centering
    \includegraphics[width=0.75\linewidth]{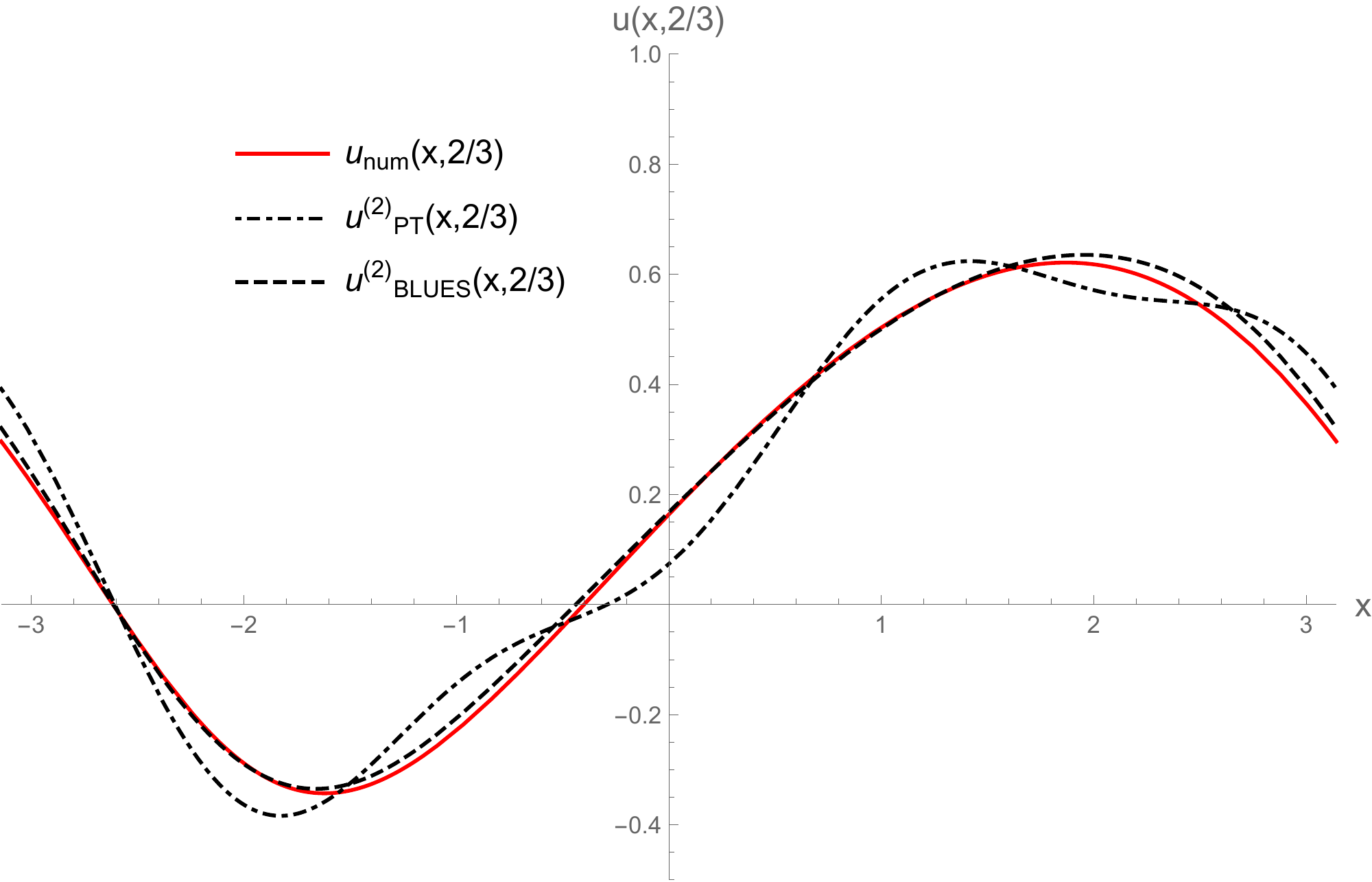}
    \caption{Periodic interface contour at time $t=2/3$. The numerical solution (red line) is compared with the $n=2$ BLUES approximant (dashed line) and the 2nd-order PT  (dot-dashed line). Parameters are $D=\alpha=1$ and $\beta=-1$.}
    \label{fig:interface_gaussian_space_PT}
\end{figure}

\begin{figure}[htp]
    \centering
    \includegraphics[width=0.75\linewidth]{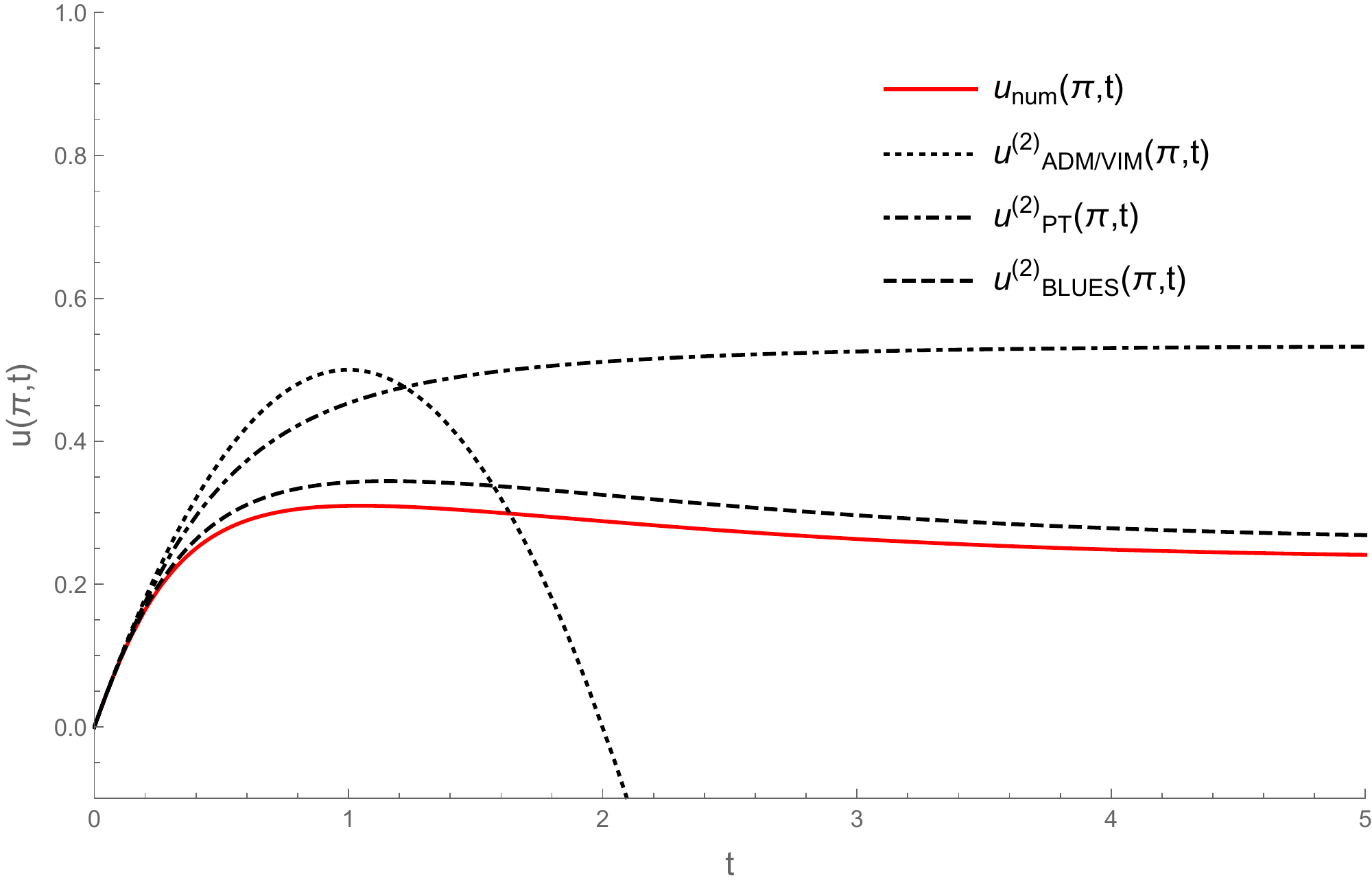}
    \caption{Periodic interface time evolution at position $x=\pi$. The numerical solution (red line) is compared with the $n=2$ BLUES approximant (dashed line), the $n=2$ approximants of ADM and VIM (dotted line), which coincide for $x=\pi$, and the $n=2$ PT (dot-dashed line). Parameters are $D=\alpha=1$ and $\beta=-1$.}
    \label{fig:interface_gaussian_time}
\end{figure}

In Fig.\ref{fig:interface_gaussian_space_PT}, we  compare the second BLUES approximant with 2nd-order PT at $t=2/3$. Finally, in Fig. \ref{fig:interface_gaussian_time} we show the various $n=2$ approximations (ADM, VIM and BLUES) for a fixed spatial coordinate $x=\pi$. We remark that the 2nd-order approximations for the ADM and VIM coincide exactly for $x=\pi$.

\begin{figure}[htp]
    \centering
    \includegraphics[width=0.85\linewidth]{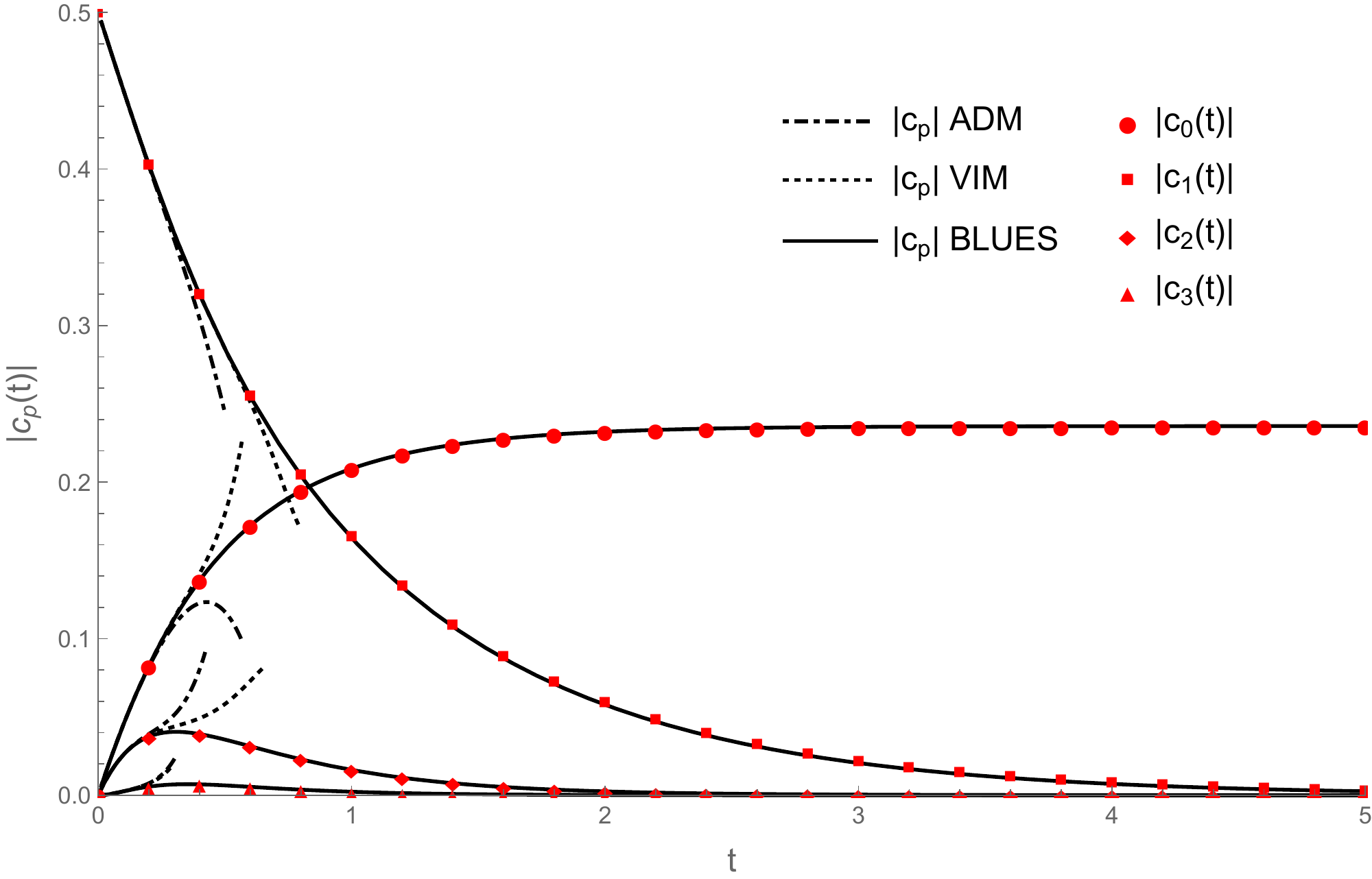}
    \caption{Time evolution of the modulus of the $p$th coefficient, for $p\in\{0,1,2,3\}$, in the Fourier series expansion of the solution of \eqref{eq:BKPZ}. The numerical solutions (red symbols) for $|c_p(t)|$ are compared with the fourth approximant of the BLUES function method (full lines),  4th-order ADM (dotted lines) and 4th-order VIM (dot-dashed lines). Parameter values are  $D=\alpha=1$ and $\beta=-1$. For ADM and VIM, the approximants are only drawn for short times, after which they  diverge uncontrollably. (For $p=3$ ADM and VIM are nearly coincident for small $t$).}
    \label{fig:Interface_fourier}
\end{figure}

From equation \eqref{eq:interface_sin_first} it is easy to see that a second harmonic is generated by both growth and convection. In further iterations higher harmonics are generated. Hence, the BLUES function method iteratively generates all harmonics as a Fourier series for which the coefficients are time-dependent. These coefficients are recursively modified by the method up to the point that they converge to their final exact value. For the function $u(x,t)$, for fixed time $t$, the complex ($c_p$) and real ($a_p$ and $b_p$) $p$th harmonic coefficients in the Fourier series  are given by
\begin{equation}
\label{eq:fourier_weights}
\begin{split}
    c_p(t) &= \frac{1}{2\pi}\int_{-\pi}^\pi dx\, u(x,t) \me^{-ipx}\\
        a_p(t) &= \frac{1}{\pi}\int\limits_{-\pi}^\pi dx\, u(x,t) \cos{px}\\
        b_p(t) &= \frac{1}{\pi}\int\limits_{-\pi}^\pi dx\, u(x,t) \sin{px},
    \end{split}
\end{equation}
with $c_p = (a_p -i b_p)/2$. In Fig.\ref{fig:Interface_fourier} the time evolution of the modulus of the coefficients $c_p(t)$ is shown for $p\in\{0,1,2,3\}$ and a comparison is made between the numerically exact values, the $n=4$ BLUES approximants and the $n=4$ ADM and VIM approximants. Note that the coefficients calculated with both the ADM and VIM diverge uncontrollably (truncated lines) as time increases while the BLUES approximants reproduce the exact coefficients almost perfectly. 
\begin{figure}[htp]
    \centering
    \includegraphics[width=0.85\linewidth]{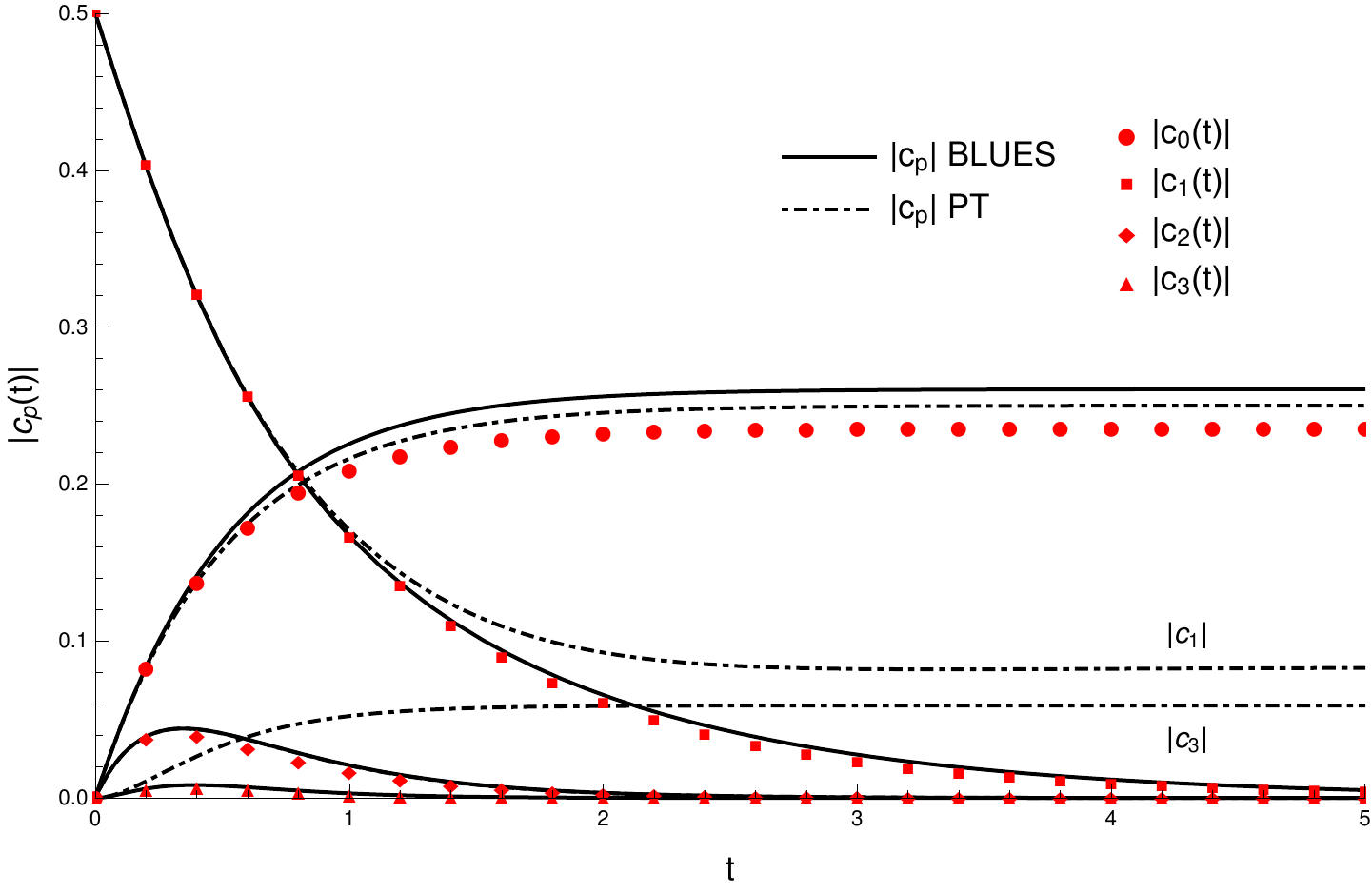}
    \caption{Time evolution of the modulus of the $p$th coefficient in the Fourier series expansion of the solution of \eqref{eq:BKPZ}. The numerical solutions (red symbols) for  $|c_p(t)|$ are compared with the second approximant of the BLUES function method (full lines) and 2nd-order PT.  Parameter values are  $D=\alpha=1$ and $\beta=-1$.}
    \label{fig:Fourier_BLUES_vs_SR}
\end{figure}

It is conspicuous that BLUES iteration progresses differently from PT. There is even a qualitative difference. For long times the asymptotic behavior of the BLUES approximants agrees with the numerically exact solution in that all the harmonics decay to zero. This is not always the case in the PT (e.g., for $|c_1(t)|$ and $|c_2(t)|$ in Fig. \ref{fig:Fourier_BLUES_vs_SR}). 

An interesting quantity is the asymptotic ``size excess" $\Delta$ of the solution as a consequence of the lateral growth correction of the interface. This is given by the long-time limit of $c_0(t)$, 
\begin{equation}
    \label{eq:Delta}
    \Delta \equiv \lim_{t\rightarrow\infty}{c_0(t)} = \frac{1}{2\pi}\lim_{t\rightarrow\infty}{\int_{-\pi}^\pi dx \,u(x,t)}.
\end{equation}
The numerically obtained precise value for the size excess is $\Delta_{\rm num} = 0.2356$, while the $n$th BLUES approximants give $\Delta^{(n=0)}_{\rm BLUES} =0$, $ \Delta^{(n=1)}_{\rm BLUES} = 0.25$, $\Delta^{(n=2)}_{\rm BLUES}= 0.2604$, $\Delta^{(n=3)}_{\rm BLUES} = 0.2421$, $\Delta^{(n=4)}_{\rm BLUES}=0.2358$. The parameter values  are $D=\alpha=1$ and $\beta=-1$. 

\section{Conclusions}
\label{sec:conclusions}
The extension of the BLUES function method to PDEs (in time and one other variable) presented here represents a significant broadening of the scope of the method. In previously reported applications to ODEs, a (co-moving) source term had to be added to the differential equation, corresponding to a physical input external to the problem and inevitably somewhat {\em ad hoc}. In contrast, in the present application to PDEs, the source term is a natural intrinsic ingredient, being the initial condition of the problem. 

In its formulation for a PDE, the BLUES iteration can be compared with various other approaches, and we have made this comparison for 4 methods: the ADM, VIM, GVIM and PT. We have observed that the BLUES iteration often provides better convergence towards the (numerically exact) solution, and offers a qualitative advantage in attaining correct asymptotic behavior for long times. This favourable position appears to result from the freedom in the method to tailor the linear operator part of the problem, so that an ``optimal" Green function becomes available (which forces correct asymptotics), and to add the remainder of the DE as a residual that contains the nonlinear operator part but also whatever remaining linear part that was not chosen to be incorporated in the linear operator. This freedom is instrumental, and requires physical insight on the side of the user before unleashing the calculation.

A physical application has been supplied, which deals with the motion and growth of an interface in a phase-separated fluid subject to shear flow. In a minimalistic model, we have combined, at the deterministic level (low-temperature approximation) the features of KPZ-like lateral growth and Burgers-like convection due to shear. A detailed study was made of a space-periodic version, with comprehensive Fourier analysis of the evolving contour. In this example, a scrutiny was made of the similarities, and differences, between (non-perturbative) BLUES and PT. This comparison has turned out, once again, to favour the former.

For the future, we envision an extension of the method to stochastic DEs for which the noise can play the role of an external source. We also consider an application to coupled DEs for which the Green function is a matrix exponential. 
In closing we note that our restriction, in this paper, to a first-order time derivative, is not necessary. We announce that the method can be readily adapted to study second-order time derivatives, by converting the problem to a system of coupled first-order PDEs. The initial conditions for the function and its derivative can then both be included, after suitable multiplication with a delta function.

\section*{Acknowledgements}
This research is supported in part by a MISTI (MIT International Science and Technology Initiatives) Global Seed Fund under project title ``Iteratively solving nonlinear growth, diffusion and convection". The authors are grateful to Mehran Kardar for hospitality and discussions at MIT (in 2019) and thank also Thorsten Emig for a pertinent question. Furthermore, we thank Timothy Halpin-Healy, Joachim Krug and Rodolfo Cuerno for encouraging comments.
\cleardoublepage
\appendix
\section{First approximant for the minimalistic interface growth model under shear}
\label{app:interfaces}
The nonlinearity in equation \eqref{eq:BKPZ} can be split up into two parts with different values for $m,n$. We first calculate the first correction to the zeroth iteration solution \eqref{eq:convolution} for the nonlinearity $\rop u = -\alpha uu_x$, which corresponds to $m=n=1$. Starting from equation \eqref{eq:lambda} and using the property \eqref{eq:hypergeo1}, the function $\Xi(x,t,s)$ reduces to
\begin{equation}
    \label{eq:reduced_Xi}
    \Xi(x,t,s) = \frac{\sqrt{\pi} \sqrt{2 D (t-s)}}{2} \frac{\Sigma^3(s)}{S^3(t,s)} x
\end{equation}
Inserting this into \eqref{eq:general_first} and keeping track of the signs results in the  correction
\begin{equation}
    \begin{split}
        u^{(1)} - u^{(0)} &= 
        \frac{\alpha}{(2\pi)^{\frac{3}{2}}}\int\limits_0^t\mathrm{d}s\frac{\mathrm{e}^{-x^2/2S^2(t,s)}}{\sqrt{2D(t-s)}  \Sigma(s)^{4}}\Xi(x,t,s)\\
        &= \frac{\alpha x}{4\sqrt{2}\pi}\int\limits_0^t\mathrm{d}s\frac{\mathrm{e}^{-x^2/2S^2(t,s)}}{\Sigma(s) S^3(t,s)}
    \end{split}
\end{equation}
By making the substitutions $\xi = S^{-1}(t,s)$, $2D(t-s) = 2 \xi^{-2} -\Sigma^2(t)$ and $\Sigma(s) = \sqrt{2}\xi^{-1} \sqrt{\xi^2 \Sigma^2(t)-1}$, the integral can be transformed into
\begin{equation}
    \label{eq:burgers_first_transformed}
    u^{(1)} - u^{(0)} = \frac{\alpha x}{4\pi D }\int\limits_{\xi_L}^{\xi_H}\mathrm{d}\xi\frac{\xi\mathrm{e}^{-x^2\xi^2/2}}{\sqrt{\Sigma^2(t)\xi^2-1}}\, ,
\end{equation}
with integration limits $\xi_L = S^{-1}(t,0) = \sqrt{2}/\Sigma(2t)$ and $\xi_H = S^{-1}(t,t) = \sqrt{2}/\Sigma(t)$. Before solving, we first proceed to calculate the first correction to the zeroth approximant \eqref{eq:convolution} for the nonlinearity $\rop u = -\beta u_x^2$, which corresponds to $m=0$, $n=2$. Starting from equation \eqref{eq:lambda} and using the property \eqref{eq:hypergeo2}, the function $\Xi(x,t,s)$ reduces to
\begin{equation}
    \label{eq:reduced_Xi2}
    \Xi(x,t,s) = \sqrt{2D(t-s)}\, \frac{\sqrt{\pi}}{2} \frac{\Sigma^3(s)}{S^3(t,s)}\left(2D(t-s) + \frac{x^2}{2}\frac{\Sigma^2(s)}{S^2(t,s)} \right)
\end{equation}
Once again inserting this into \eqref{eq:general_first} and keeping track of the signs results in the correction
\begin{equation}
    \begin{split}
        u^{(1)} - u^{(0)} &= -\frac{\beta}{4\pi\sqrt{2}}\int\limits_0^t\mathrm{d}s\frac{\mathrm{e}^{-x^2/2S^2(t,s)}}{\Sigma^3(s) S^3(t,s)}\left(2D(t-s)+\frac{\Sigma^2(s)x^2}{2S^2(t,s)}\right)
    \end{split}
\end{equation}
By making the same substitutions as before the integral can be transformed into
\begin{equation}
    \label{eq:kpz_first_transformed}
    u^{(1)} - u^{(0)} = \frac{-\beta}{8\pi D }\int\limits_{\xi_L}^{\xi_H}\mathrm{d}\xi\frac{\xi\mathrm{e}^{-x^2\xi^2/2}}{\sqrt{\Sigma^2(t)\xi^2-1}} \left(x^2 \xi^2 -1 +\frac{1}{\Sigma^2(t) \xi^2 -1}\right)
\end{equation}
Finally, combining equations \eqref{eq:burgers_first_transformed} and \eqref{eq:kpz_first_transformed}, the first correction to the zeroth approximant becomes
\begin{equation}
    \label{eq:total_first_transformed}
    u^{(1)} - u^{(0)} = \frac{1}{4\pi D }\int\limits_{\xi_L}^{\xi_H}\mathrm{d}\xi\frac{\xi\mathrm{e}^{-x^2\xi^2/2}}{\sqrt{\Sigma^2(t)\xi^2-1}} \left(\frac{\beta}{2} + \alpha x -\frac{\beta x^2\xi^2}{2} -\frac{\beta}{2\left(\Sigma^2(t)\xi^2-1\right)}\right)\, ,
\end{equation}
which can easily be solved and subsequently simplified by noticing that $2\Sigma^2(t) - \Sigma^2(2 t) = \sigma^2$ to give the following expression for the correction in first iteration to the zeroth approximant of \eqref{eq:BKPZ}
\begin{equation}
    \label{eq:total_first_solved}
    \begin{split}
         u^{(1)} - u^{(0)} &= \frac{\beta}{4\pi D}\left[\frac{ \mathrm{e}^{-x^2/\Sigma^2(t)}}{\Sigma^2(t)}-\frac{ \mathrm{e}^{-x^2/\Sigma^2(2t)}}{\Sigma(2t)\sigma}\right] \\
         &+\frac{\alpha}{4D\sqrt{2\pi}}\left[\frac{\mathrm{e}^{-x^2/2\Sigma^2(t)}}{\Sigma(t)}\left(\erf{\left(\frac{x}{\sqrt{2}\Sigma(t)}\right)} -\erf{\left(\frac{\sigma x}{\sqrt{2}\Sigma(t)\Sigma(2 t)}\right)}\right)\right]\, .
    \end{split}
\end{equation}
This can now be rearranged to yield equation \eqref{eq:interface_first_solved}.

\section{Fourier coefficients for the space-periodic interface contour}
\label{app:fourier}
\begin{figure}[htp]
    \centering
    \includegraphics[width=0.85\linewidth]{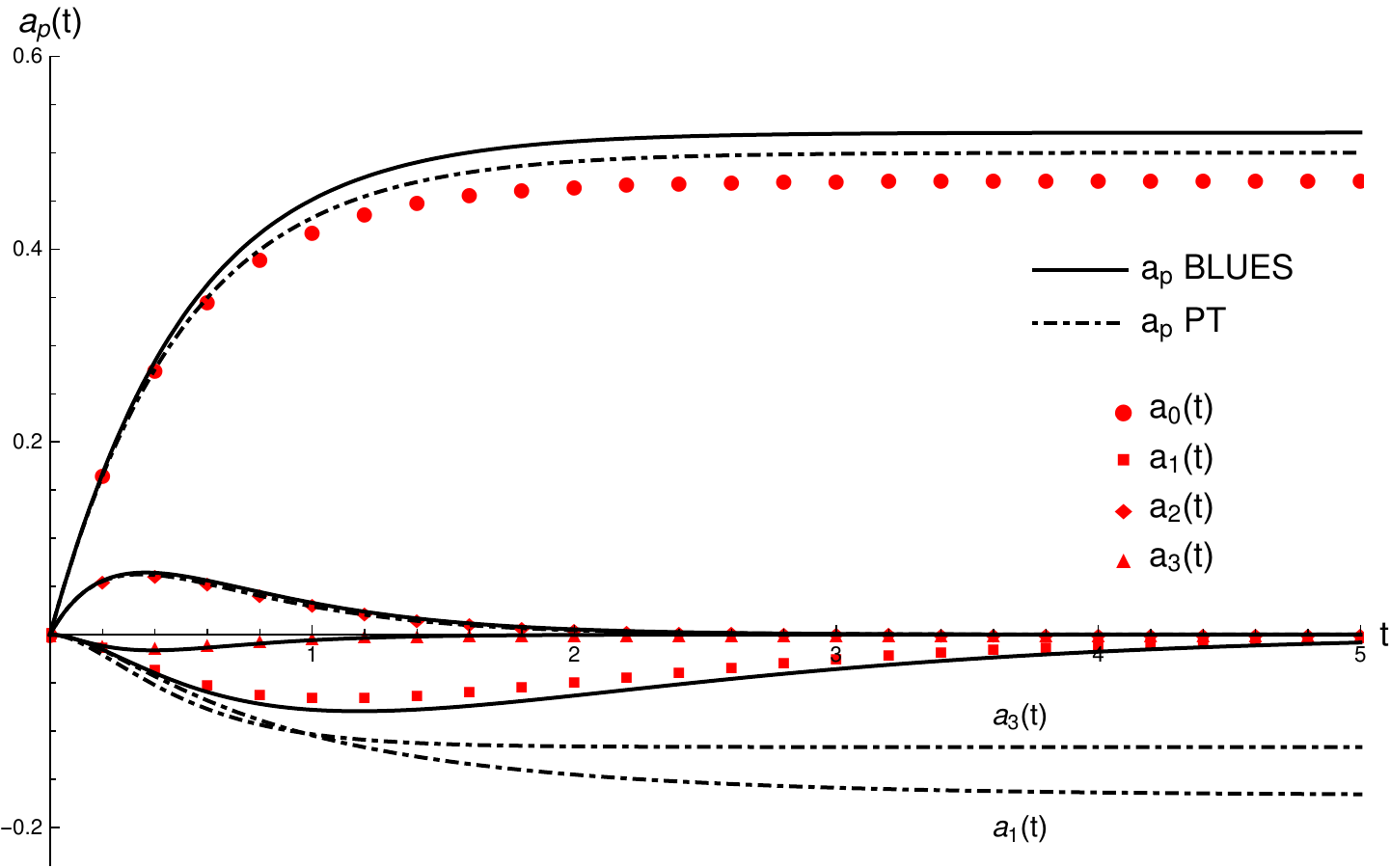}
    \caption{Time evolution of the  coefficients $a_p(t)$ of the cosine harmonics in the Fourier series expansion of the solution of \eqref{eq:BKPZ}. The numerical solutions (red symbols) for  $a_p(t)$ are compared with the second approximants of the BLUES function method (full lines) and 2nd-order PT.  Parameter values are  $D=\alpha=1$ and $\beta=-1$.}
    \label{fig:Interface_FourierCosPlot}
\end{figure}

\begin{figure}[htp]
    \centering
    \includegraphics[width=0.85\linewidth]{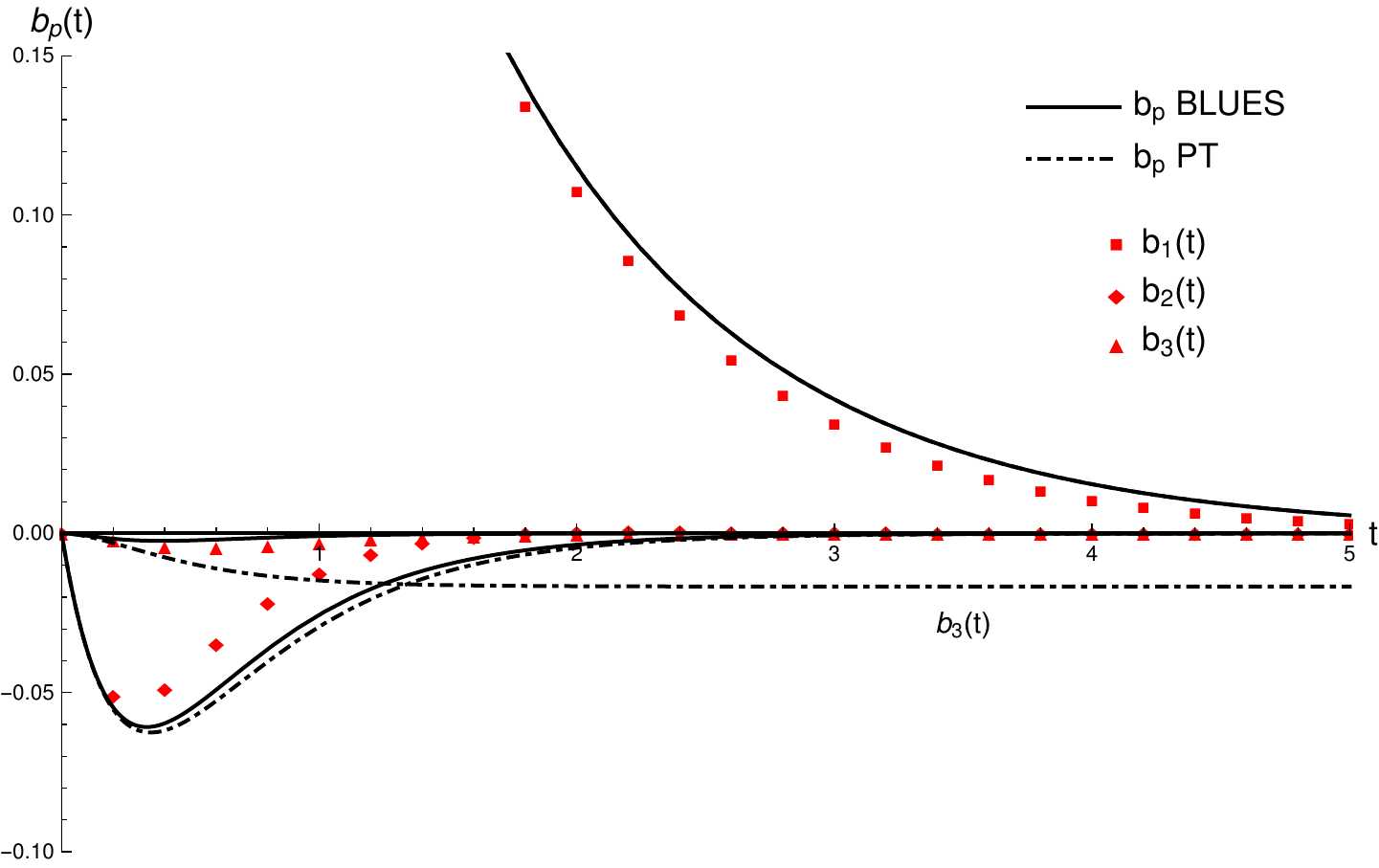}
    \caption{Time evolution of the  coefficients $b_p(t)$ of the sine harmonics in the Fourier series expansion of the solution of \eqref{eq:BKPZ}. The numerical solutions (red symbols) for  $b_p(t)$ are compared with the second approximants of the BLUES function method (full lines) and 2nd-order PT.  Parameter values are  $D=\alpha=1$ and $\beta=-1$.}
    \label{fig:Interface_FourierSinPlot}
\end{figure}

In this Appendix we discuss in detail the Fourier coefficients of various harmonics that are generated by the BLUES iteration at the level of the second approximant ($n=2$) for the problem of the time evolution of the periodic interface contour and compare them with their counterparts in 2nd-order PT. We first present, for $p\in\{0,1,2,3\}$, the real $p$th coefficients calculated by both methods and then discuss them with the aid of two figures, \ref{fig:Interface_FourierCosPlot} and \ref{fig:Interface_FourierSinPlot}.

For the $a_n$ we obtain:
\begin{itemize}
    \item $a_0$ BLUES:
    \begin{equation}
        a_0(t) = -\frac{\beta(1-e^{-2 D t})}{2D}  -\frac{ \beta   \left (\alpha ^2+\beta ^2\right)\left(1-e^{-2 D t}\right)^3 \left(3+e^{-2 D t}\right)}{96 D^3}
    \end{equation}
    \item $a_0$ PT:
    \begin{equation}
        a_0(t) = -\frac{\beta  (1-e^{-2 D t})}{2D}
    \end{equation}
       \item $a_1$ BLUES: 
    \begin{equation}
        a_1(t) = -\frac{\alpha \beta\left(e^{-Dt} + 2e^{-3 Dt} -3 e^{-5 Dt}\right)}{32D^2} + \frac{\alpha \beta e^{-Dt}\,t}{4D}
    \end{equation}
    \item $a_1$ PT:
    \begin{equation}
        a_1(t) = \frac{\alpha\beta (16 - 15 e^{-Dt} - 10 e^{-3 Dt} + 9 e^{-5 Dt})}{96D^2}
    \end{equation}
    \item $a_2$ BLUES
    \begin{equation}
        a_2(t) = -\frac{\beta (e^{-2 D t} - e^{-4 Dt})}{4 D} - \frac{\alpha^2\beta(e^{-2 Dt} - e^{-6Dt})}{16D^3} + \frac{\alpha^2\beta e^{-4 Dt} t}{4D^2}
    \end{equation}
    \item $a_2$ PT:
    \begin{equation}
        a_2(t) = -\frac{\beta  (e^{-2 D t}-e^{-4 Dt})}{4D}
    \end{equation}
    \item $a_3$ BLUES:
    \begin{equation}
        a_3(t) =\frac{7 \alpha\beta (1-e^{-2 Dt})^2 (2 e^{-3 Dt} + e^{-5 Dt})}{96 D^2}
    \end{equation}
    \item $a_3$ PT:
    \begin{equation}
        a_3(t) = \frac{7 \alpha\beta (2- 5 e^{-3 D t} +  3 e^{-5 D t} )}{120D^2}
    \end{equation}
    \item $a_4$ BLUES:
    \begin{equation}
        a_4(t) = \frac{\beta(\beta^2-2\alpha^2) (1-e^{-2 D t})^3 (10 e^{-4 Dt} + 6 e^{-6 Dt} + 3 e^{-8 Dt} + e^{-10Dt})}{960D^3}
    \end{equation}
\end{itemize}
For the  $b_n$ we obtain:
\begin{itemize}
    \item $b_0=0$  BLUES and PT.
    \item $b_1$ BLUES:
    \begin{equation}
        b_1(t) = e^{-Dt} - \frac{(\alpha^2 + 4 \beta^2)(e^{-Dt}- 2 e^{-3 Dt} + e^{-5 Dt} )}{32 D^2}
    \end{equation}
    \item $b_1$ PT:
    \begin{equation}
        b_1(t) = e^{-Dt} - \frac{(\alpha^2 + 4 \beta^2)(e^{-Dt}- 2 e^{-3 Dt} + e^{-5 Dt} )}{32 D^2}
    \end{equation}
    \item $b_2$ BLUES:
    \begin{equation}
        b_2(t) = -\frac{\alpha (e^{-2 Dt} - e^{-4 Dt})}{4 D} + \frac{\alpha\beta^2 (e^{-2 Dt} - e^{-6 Dt})}{16D^3} - \frac{\alpha\beta^2 e^{-4 Dt}\,t}{4 D^2}
    \end{equation}
    \item $b_2$ PT:
    \begin{equation}
        b_2(t) = -\frac{\alpha (e^{-2Dt} - e^{-4Dt})}{4 D}
    \end{equation}
    \item $b_3$ BLUES:
    \begin{equation}
        b_3(t) = \frac{(3\alpha^2 - 4\beta^2) (1-e^{-2 Dt})^2 (2 e^{-3 Dt} + e^{-5 Dt})}{96 D^2}
    \end{equation}
    \item $b_3$ PT:
    \begin{equation}
        b_3(t) = \frac{(3\alpha^2 - 4\beta^2) (2- 5 e^{-3 Dt}+3 e^{-5 Dt})}{120D^2}
    \end{equation}
    \item $b_4$ BLUES:
    \begin{equation}
        b_4(t) = -\frac{\alpha(\alpha^2 - 5\beta^2) (1-e^{-2 Dt})^3 (10 e^{-4 Dt} + 6 e^{-6 Dt} + 3 e^{-8 Dt} + e^{-10Dt})}{1920D^3}
    \end{equation}
\end{itemize}

Note that in the second approximant for $a_0$ terms of order $\alpha^2\beta$ and $\beta^3$ are generated, which are absent in 2nd-order PT. Als note that $a_1$ (first harmonic) and $a_3$ (third harmonic) are both proportional to $\alpha\beta$, as in PT. Importantly, in the BLUES function method $a_1(t)$ and $a_3(t)$ tend to zero for long times, in agreement with the numerical solution, whereas the 2nd-order PT expressions tend to non-zero constants (see also Fig. \ref{fig:Interface_FourierCosPlot}). In this respect the BLUES iteration is qualitatively superior. 
The coefficient $a_2(t)$ (second harmonic) has a first order in $\beta$ contribution which is the same in both methods, and an additional $\alpha^2\beta$ contribution in the second BLUES approximant.  In both methods the result is very close to the numerical solution (see Fig. \ref{fig:Interface_FourierCosPlot}).
Note that $a_4(t)$ (fourth harmonic) is generated in 2nd-iteration BLUES but is absent in 2nd-order PT. This is a consequence of the fact that the BLUES function method is non-perturbative and already generates higher harmonics in a lower iteration than the perturbation series.

As for the $b_n(t)$, the coefficient $b_1(t)$ (first harmonic reflecting the initial condition) contains the zeroth approximant, which is (of course) the same in both methods.  Moreover, the entire expressions for $b_1(t)$ coincide in 2nd-iteration BLUES and 2nd-order PT (see also Fig. \ref{fig:Interface_FourierSinPlot}). The coefficient $b_2(t)$ (second harmonic) has a first order in $\alpha$ contribution which is the same in both methods, and an additional $\alpha\beta^2$ contribution in the 2nd BLUES approximant.  In both methods the result is nearly the same but both are somewhat off of the numerical solution (see Fig. \ref{fig:Interface_FourierSinPlot}). Importantly, in the BLUES function method $b_3(t)$ (third harmonic) tends to zero for long times, in agreement with the numerical solution, whereas the 2nd-order PT expression tends to a non-zero constants (see also Fig. \ref{fig:Interface_FourierSinPlot}). In this respect the BLUES iteration is again qualitatively superior. Finally, $b_4(t)$ (fourth harmonic) is present in BLUES but is obviously absent in 2nd-order PT because it is of third order.

\bibliography{BLUES.bib}
\end{document}